\documentclass[aps,twocolumn,superscriptaddress,floatfix,prb]{revtex4-2}

\usepackage{lipsum}
\usepackage{graphicx}
\usepackage{amsmath,amssymb}
\usepackage{braket}
\usepackage{mathtools}
\usepackage{hyperref}
\usepackage{multirow}
\usepackage{color}
\usepackage[normalem]{ulem}
\usepackage{amsfonts}
\usepackage{float}
\usepackage{pdfpages} 
\makeatletter
\usepackage{subfigure}
\setcitestyle{super}

\def\beq{\begin{equation}}
\def\eeq{\end{equation}}
\def\bea{\begin{eqnarray}}
\def\eea{\end{eqnarray}}

\AtBeginDocument{\let\LS@rot\@undefined}
\makeatother

\begin{document}


\title{Onset of many-body quantum chaos due to breaking integrability}

\author{Vir B. Bulchandani}
\affiliation{Princeton Center for Theoretical Science, Princeton University, Princeton, NJ 08544, USA}
\affiliation{Department of Physics, Princeton University, Princeton, NJ, 08544, USA}
\author{David A. Huse}
\affiliation{Department of Physics, Princeton University, Princeton, NJ, 08544, USA}
\affiliation{Institute for Advanced Study, Princeton, NJ, 08540, USA}
\author{Sarang Gopalakrishnan}
\affiliation{Department of Physics, Pennsylvania State University, University Park PA 16802, USA}

\begin{abstract}

Integrable quantum systems of finite size are generically robust against weak enough integrability-breaking perturbations, but become quantum chaotic and thermalizing if the integrability-breaking is strong enough. 
We argue that the onset of quantum chaos can be described as a Fock-space delocalization process, with the eigenstates of the integrable system being taken as the ``Fock states''.  The integrability-breaking perturbation introduces 
hopping in this Fock space, and chaos sets in when this hopping delocalizes the many-body eigenstates in this space. 
Depending on the range of the dominant Fock-space hopping, delocalization can occur either through a crossover, or via a transition that becomes sharp in the appropriate large-system {\it dynamic limit}.
In either case, the perturbation strength at the onset of chaos scales to zero in the usual thermodynamic limit, with a size-dependence that we estimate analytically and compute numerically for a few specific models. 
We also identify two intermediate finite-size-dependent regimes:  There is generally an intermediate nonchaotic regime in which integrability is broken strongly enough to produce some system-wide many-body resonances but not enough to thermalize the system. In spatially extended systems (but not in quantum dots) there is also a crossover or transition between chaotic regimes where the ratio of the system size to the mean free path of the quasiparticles of the integrable system is small versus large compared to unity.

\end{abstract}

\maketitle

\section{Introduction}
\subsection{Thresholds to many-body chaos}

For much of the 20th century, there was a widespread belief that arbitrarily weak perturbations of classical integrable systems lead to chaos. It was therefore a great surprise when Kolmogorov, Arnol'd and Moser (KAM) showed that sufficiently weak perturbations of classical integrable systems preserve integrability within an appreciable fraction of phase space. In particular, they found that a perturbation whose strength scales with a small parameter $\epsilon$ preserves integrable dynamics in a fraction $1-\mathcal{O}(\epsilon^{1/2})$ of the phase space, that is characterized by the persistence of so-called KAM tori\cite{poschel}.

At a first encounter, this result sits uneasily with the dogma of ergodicity on energy surfaces that underpins much of statistical physics. If integrability truly is robust to perturbations, how can ergodicity arise generically in large systems? It is believed that the answer to this question lies in the finite-size scaling of the threshold perturbation strength $\epsilon_c(N)$ at which KAM tori break down, where $N$ denotes the number of degrees of freedom of the system. In particular, conventional wisdom states that $\epsilon_c(N) \to 0$ as $N \to \infty$.

Unfortunately, the problem of analysing perturbed integrable systems at large $N$ is so difficult that there is little analytical understanding of the threshold $\epsilon_c(N)$, even for classical systems\cite{Wayne,broer2010kam}. The original KAM approach is guaranteed to converge for a range of perturbation strengths\cite{Wayne}
\begin{equation}
\epsilon < \frac{C}{(N!)^{31}}
\end{equation}
which is certainly not inconsistent with the ergodic hypothesis. To our knowledge, the best improvement of this result is to perturbation strengths
\begin{equation}
\label{eq:Wayne}
\epsilon < \frac{C'}{N^{160}}
\end{equation}
for systems with short-range interactions, but there is no expectation that this result is optimal\cite{Wayne}.

This paper studies the quantum mechanical analogue of the problem described above. Specifically, we propose a theoretical model for the critical perturbation strength $\epsilon_c(N)$ at which perturbed \emph{quantum} integrable systems give way to chaos. On the one hand, this task is complicated by the fact that there is no analogue of the KAM theory for quantum systems, which partly reflects some technical difficulties in robustly defining integrability and chaos for quantum systems\cite{Caux_2011,berry1987bakerian}. On the other hand, the problem of finding $\epsilon_c(N)$ is in some ways \emph{more} tractable for quantum systems, first because a relatively simple theory of ``Fock-space delocalization'' seems to describe the onset of chaos in perturbed non-interacting quantum systems\cite{AGKL}, and second because it is possible to simulate numerically the entire spectrum of small systems of spins or fermions, which is never the case for continuous classical systems. We now summarize the physical motivation for understanding quantum integrability breaking, and the theoretical model that we will use to describe it.

\subsection{Quantum integrability breaking: overview}

There are numerous many-body quantum systems of interest that are in some sense nearly integrable. This can be because the system is very close to an integrable system but has a weak integrability-breaking perturbation. Or it can be because the system is near a ground state that features an emergent integrability, even if the same system's dynamics are far from integrable at higher energies. Although the study of quantum integrability remained largely theoretical for decades, present-day experimental capabilities for the preparation and control of many-body quantum systems are sufficiently advanced that several recent experiments have observed unambiguous signatures of integrable dynamics and transport in such systems\cite{cradle,PhysRevX.8.021030,AtomChip,Scheie_2021,WeissRecent,BlochKPZ}.

At the same time, it is well known that if one perturbs a nonlocalized quantum integrable system, such as a spin chain, with an integrability-breaking perturbation of any non-zero strength, the system thermalizes in the conventional thermodynamic limit, in which all local couplings are kept constant as the system size is taken to infinity~\cite{millis,santos2004integrability, PhysRevLett.125.180605}. This can be true even if the perturbation acts only locally at a single
point~\cite{santos2004integrability, PhysRevE.89.062110, PhysRevB.98.235128, bastianello2019lack, PhysRevLett.125.070605}. Many studies of systems with weakly-broken integrability are numerical or experimental studies of finite systems, with substantial finite size effects.  Thus we are not only interested in infinite systems.  To better understand the observable behavior in finite size systems, it is helpful to consider large system limits that are different from the standard thermodynamic limit.

The distinction between integrability and chaos is not an equilibrium thermodynamic distinction; instead, it is a distinction in the system's {\it dynamics}. Thus in order to understand more about how an integrability-breaking perturbation to an integrable system leads to chaos, particularly in finite-size systems, below we focus our attention on the range of perturbation strengths that is relevant to the onset of chaos.  This requires scaling down the integrability-breaking perturbation as one takes the large-system limit so that the system remains close to the onset of chaos for all sizes---we call this limiting procedure a \emph{dynamic limit}. In such a dynamic limit, the integrability-breaking perturbation has no effect on the thermodynamics of the system but qualitatively alters its dynamics. A textbook example of this is the very weak interactions in a dilute gas that are crucial for its equilibration but do not affect its equation of state. A single system may have multiple dynamic limits, depending on the particular dynamical phenomenon of interest.

A pioneering investigation of such ``dynamic limits'' was performed by Altshuler, Gefen, Kamenev and Levitov\cite{AGKL} (AGKL), who examined the integrability-breaking ``many-body delocalization'' transition that occurs when weak interactions are added to a quantum dot containing many noninteracting electrons. In that work, as well as subsequent work on various other many-body localized (MBL) models\cite{Burin,Ponte,TikhonovMirlin2018,GopalakrishnanHuse}, it was argued that sharp phase transitions can happen in the dynamic limit: i.e., transitions between ``phases'' with identical thermodynamic properties, but sharply different dynamics in the long-time limit. (Note that the one-dimensional MBL transition~\cite{BAA} is also such a dynamical transition, albeit one that happens in the conventional thermodynamic limit.) 
However, there has been relatively little work from this perspective on integrability-breaking in conventionally integrable systems, which are solvable via the Bethe ansatz and have stable ballistically propagating quasiparticles. 
Much of the work on integrability-breaking in Bethe-solvable systems has addressed transport, quench dynamics, etc. in the conventional thermodynamic limit~\cite{Pretherm,hardrods, PhysRevX.8.021030, PhysRevLett.120.070603, PhysRevB.101.180302, PhysRevB.102.161110, PhysRevLett.126.090602, PhysRevLett.127.130601, PhysRevE.103.042121, bouchoule2020effect, bastianello2021hydrodynamics}. In this limit, the system is chaotic on timescales longer than the mean free time of the quasiparticles. Much less is known about the behavior in the \emph{dynamic limit} relevant for the onset of chaos: even the basic question of whether chaos sets in at a perturbation strength that is polynomially or exponentially weak in the system size remains unsettled, with numerical evidence for both scalings~\cite{SpeckChaos,PolkovnikovSels,Modak,Budapest,fogarty2021probing}; as we will argue below, both behaviors do occur, depending on the specific model under consideration.

\subsection{Quantum integrability breaking: a theoretical model}

In the present work, we study finite-size systems and these dynamic limits systematically by exploiting an analogy between perturbed integrable systems and Fock-space delocalization in quantum dots (as discussed by AGKL). Each eigenstate of an integrable system can be uniquely labeled by the values of extensively many local conserved quantities in that eigenstate, or equivalently by an extensive set of quantum numbers. Heuristically, the existence of such labeling schemes suggests a natural metric on state space, according to which states with ``similar'' labels are ``near'' one another. For example, in the AGKL analysis of a quantum dot, eigenstates can be labeled by bit-strings corresponding to whether each single-particle orbital is occupied or not. These bit-strings can be pictured as the vertices of a hypercube, and the natural metric is the Hamming distance between bit-strings.

The key observation that justifies the reduction of \emph{local} integrable systems to effectively \emph{zero-dimensional} Fock-space localization is that the thermalization rate at the onset of chaos is close to the \emph{Heisenberg} energy (i.e., exponentially small in the system's volume), and is thus much slower than the time it takes ballistic quasiparticles to traverse the system. Thus, on timescales relevant to the onset of chaos a finite weakly-perturbed integrable system with nonlocalized quasiparticles is essentially a zero-dimensional quantum dot. This observation places such local integrable systems (like the spin-$1/2$ Heisenberg chain) and all-to-all integrable systems (like Gaudin magnets) on an equal footing, and allows us to identify two possible main scenarios for the onset of chaos:

\begin{enumerate}
\item The integrability-breaking perturbation induces sufficiently short-range hopping in Fock space.  When this scenario holds, the onset of chaos is a \emph{transition} in the appropriate large-system dynamic limit, and the critical perturbation strength is polynomially small in the system size $N$. This scenario occurs for example in the perturbed non-interacting quantum dot considered by AGKL: in that case the interaction can only rearrange four occupation numbers at once, so it is strictly short-range on the Fock-space hypercube. But this short-range scenario can also apply when the Fock-space hopping is not strictly short-range, provided that the short-range hopping is strong enough relative to the long-range hopping.
\item The hopping is instead sufficiently long-range in Fock space that highly nonlocal processes dominate near the onset of chaos. We present evidence that this scenario occurs in XXZ spin chains subject to a class of perturbations that mix parity sectors and thus break the mapping to Jordan-Wigner fermions. When this scenario holds, the onset of chaos remains a \emph{crossover} even in the large-system dynamic limit, and the perturbation strength at this crossover is exponentially small in $N$.
\end{enumerate}

We emphasize that this observation would not be valid for many-body localized systems with short-range couplings, which instead require a theoretical treatment that accounts for the distribution of resonances in real space\cite{BAA,imbrie2016many}. However, for systems with non-localized quasiparticles, this observation implies that the physics of integrability breaking can be understood directly from the matrix elements of the perturbation in the unperturbed basis, which define the amplitudes for hopping in Fock space and therefore determine whether there is a transition or a crossover to chaos.

Unfortunately, even though these matrix elements (the so-called ``form factors'' of local operators~\cite{gohmann2017thermal, de2018particle, granet2021out}) can be efficiently computed in some specific models, their general statistical properties have not been explored in detail (though see Ref.~\cite{PhysRevE.100.062134}). Therefore, our exploration of the onset of chaos will be largely limited to small systems that are amenable to exact diagonalization (though we will briefly discuss one case where larger system sizes are attainable, see Sec. \ref{SecIV}.) Using Bethe ansatz technology to extend our analysis to larger system sizes is a desirable goal for future work.

For the same reason, we do not consider perturbations of integrable systems with infinite-dimensional Hilbert spaces in this work, such as conformal field theories and integrable bosonic theories like the Lieb-Liniger model. In the former case, we note that the large degeneracy of the unperturbed integrable system implies that integrability is already broken by processes that occur at first order in perturbation theory
, as discussed recently in the literature\cite{CFTchaos}, so that a detailed analysis of the
high-order processes leading to Fock-space delocalization is not necessary for understanding the onset of chaos. For perturbed integrable field theories with non-degenerate spectra (for example, perturbed, non-degenerate free bosons) we expect the analysis presented in this paper to hold at any finite temperature, since even for an infinite-dimensional Hilbert space, the onset of chaos near a microcanonical energy shell is constrained by the Fock-space hopping range of integrability-breaking perturbations.

The rest of this paper is organized as follows. In the remainder of this section we present an overview of the crossovers and/or transitions that separate the  integrable and chaotic regimes at finite size and in the appropriate dynamic limits.  (To the best of our knowledge these crossovers have not been discussed this thoroughly elsewhere.)  In Sec. \ref{SecII} we summarize the Fock-space delocalization picture introduced by AGKL, and discuss its relevance to perturbed integrable systems. In Sec. \ref{SecIII} we present a numerical study of perturbed Gaudin magnets, and in Sec. \ref{SecIV} we turn to quantum spin chains. Finally, we identify the key open questions raised by our analysis.

\subsection{From integrability to chaos: a summary}

\begin{figure}[tb]
\begin{center}
\includegraphics[width=0.47\textwidth]{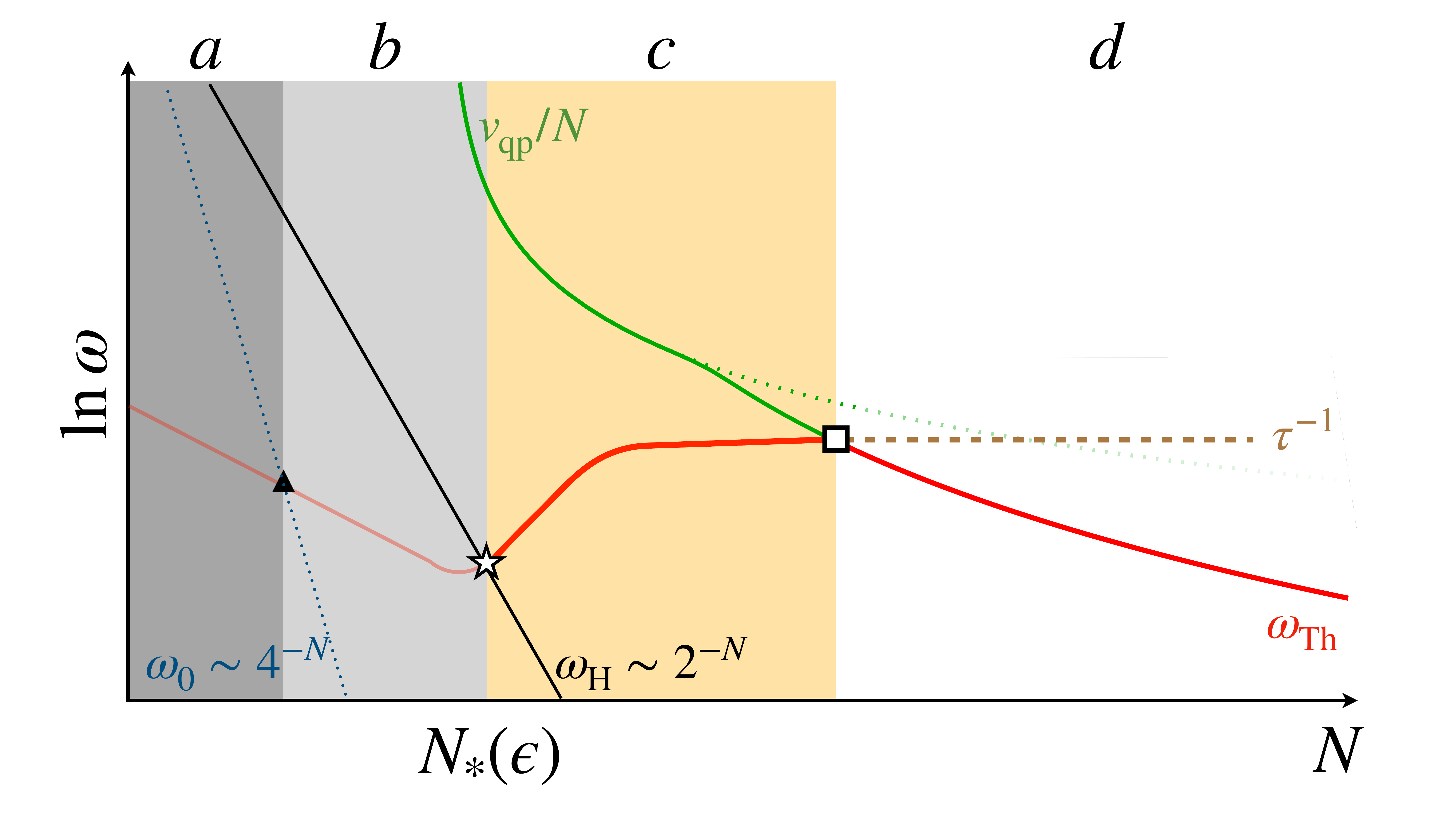}
\caption{Schematic evolution of some of the lowest characteristic energy scales of a nearly integrable spatially extended one-dimensional system with system size $N$ at fixed small integrability-breaking perturbation $\epsilon$. In the chaotic regimes $(c), (d)$, the Thouless energy $\omega_{\mathrm{Th}}$ is indicated by a thick red line.  In the localized regimes $(a), (b)$ the thin red line is the crossover scale below which most pairs of near-in-energy eigenstates are subject to substantial level repulsion. This schematic is sketched for the scenario in which delocalization is a crossover, so delocalization occurs at lowest order in $\epsilon$; in the transition scenario this thin red line would curve upwards. 
From left to right, crossovers occur when the red lines cross the smallest gap in a typical sample ($\omega_0$), the typical level spacing ($\omega_H$), and the characteristic rate for quasiparticles to traverse the system ($v_{\mathrm{qp}}/N$). In some other systems the boundary between $(b)$ and $(c)$ can sharpen up to a Fock-space delocalization transition in the large-system dynamic limit. 
In regime~$(c)$ the Thouless energy is set by the slowest-decaying spatially uniform relaxation rate ($\tau^{-1}$) , while in regime~$(d)$ it is set by the slowest spatial transport of the conserved densities that remain conserved in the presence of the perturbation.}
\label{schematic}
\end{center}
\end{figure}

A general summary of some of the features of the change from integrability to chaos is as follows:  We consider systems with $N$ degrees of freedom (usually spin-1/2 spins) and Hamiltonian
\begin{equation}\label{generalh}
\hat H=\hat H_0 + \epsilon \hat V~.    
\end{equation}
$\hat H_0$ is integrable, where this includes noninteracting particles. $\hat V$ breaks the integrability. Thus for $\epsilon\neq 0$ the full Hamiltonian $\hat H$ breaks the local conservation laws of $\hat H_0$, except for energy and (possibly) particle number. Both $\hat H_0$ and $\hat V$ are scaled with $N$ to be extensive, so that each defines a proper thermodynamic Hamiltonian in its own right. We define our ``Fock space'' as the many-body eigenstates of $\hat H_0$. $\hat V$ produces hopping in that Fock space.

We now consider the crossovers that take place as one increases $N$ while keeping $\epsilon$ fixed and much smaller than the characteristic energy scales of $\hat H_0$ (Fig.~\ref{schematic}). (We can equivalently consider increasing $\epsilon$ at fixed $N$, and will use both perspectives interchangeably below.)
In this summary, we focus, for concreteness, on the many-body level statistics.  For completeness, we also include systems where $\hat H_0$ is MBL in this summary, although we do not report any new work about MBL systems in this paper.

The first regime~$(a)$ occurs for small $N$ (note that we are taking $\epsilon$ small enough that even in this regime we can assume $N \gg 1$).  Here, the level spacing of the system is large enough that \emph{any} two adjacent levels of $\hat H_0$ are too far apart in energy for the perturbation to hybridize them.
{\it All} of the eigenstates of $\hat H$ remain localized in Fock space close to eigenstates of $\hat H_0$, so there are no system-wide many-body resonances. Here and throughout the remainder of the paper, we define a ``many-body resonance'' as a resonance between two eigenstates of $H_0$ that differ from each other in extensively many quantum numbers.
(For discussions of many-body resonances in the contexts of MBL and/or Floquet heating, see Refs.~\onlinecite{PhysRevB.93.155132, PhysRevLett.114.140401, PhysRevB.95.014112,PhysRevB.92.104202, crowley2020constructive, PhysRevB.104.184203, villalonga2020eigenstates, morningstar2021avalanches, new_marko_z}.)  If $\hat H_0$ is drawn from an ensemble of systems, in this regime $(a)$ there are no such resonances 
in almost all samples. However, some rare samples will have accidental near-degeneracies in the spectrum of $\hat H_0$ and these may be lifted by the perturbation.

The second regime~$(b)$ of weak integrability-breaking is where $\epsilon$ is large enough to produce exponentially many (in $N$) eigenstates that are system-wide many-body resonances.  These resonant eigenstates are linear combinations between Fock states that differ extensively.  These states are subject to energy-level repulsion, thus producing weak deviations from Poisson level statistics.  But in this second regime almost all eigenstates remain nonresonant and localized near eigenstates of $\hat H_0$, and the system thus remains localized in Fock space and nonchaotic: some observable information about the initial state remains exactly conserved~\cite{Glimmers}. The crossover between these first two regimes can be detected by measures akin to the ``adiabatic gauge potential''~\cite{PolkovnikovSels, PhysRevB.104.L201117} that are sensitive to the presence of these rare resonances.  The value of $\epsilon$ at this first crossover can range from exponentially small in $N$ for some interacting integrable systems, to power-law small in $N$ if $\hat H_0$ is noninteracting delocalized particles, to of order one (but still small) if $\hat H_0$ is many-body localized in one dimension and both $\hat H_0$ and $\hat V$ are sufficiently short-ranged in real space.

The third regime~$(c)$ of weak integrability-breaking is one of quantum chaos and thermalization. This regime sets in when the {scale for level repulsion between distant-in-Fock-space eigenstates} becomes comparable to the typical level spacing: thus, \emph{typical} many-body eigenstates become involved in resonances and Fock-space localization breaks down. In this regime, a new characteristic energy scale emerges---the Thouless energy, $\omega_{\mathrm{Th}}$: 
For energy differences $\omega < \omega_{\mathrm{Th}}$ the level statistics is that of a random matrix. At the boundary between regimes~$(b)$ and $(c)$, the Thouless energy becomes comparable to the typical level spacing, as long-Fock-space-range, slow relaxation processes become on-shell, destabilizing Fock-space localization. 
As one proceeds deeper into regime~$(c)$, faster decay channels (corresponding to shorter-range processes in Fock space) go on-shell and thus open up, and $\omega_{\mathrm{Th}}$ increases. We emphasize that throughout this regime, the slowest relaxation mode of the system corresponds to \emph{Fock-space} delocalization, and the physical (real-space) dimensionality of the system is still irrelevant.

Deep in regime~$(c)$, the Thouless energy scales as $\epsilon^2$ by Fermi's Golden Rule, but its scaling near the onset of Fock-space delocalization can be anomalous. At a fixed $\epsilon$, there is some characteristic system size $N^*(\epsilon)$ (given by inverting $\epsilon_c(N)$) at which the system enters regime~$(c)$. At the onset of regime~$(c)$, the Thouless energy is $\omega_{\mathrm{Th}} \sim 2^{-N^*(\epsilon)}$. Thus if $ 2^{-N/2} \ll \epsilon_c$ in the limit of large $N$ (as appears to be the case in all the models we study), the Thouless energy at the onset of chaos is parametrically smaller in $\epsilon$ than $\epsilon^2$. To match the scaling of $\omega_{\mathrm{Th}}$ at the onset of regime~$(c)$ with the behavior deep inside regime~$(c)$, $\omega_{\mathrm{Th}}$ must increase by a parametrically large amount in regime~$(c)$, as illustrated in Fig.~\ref{schematic}.

Finally, for systems that are spatially extended and such that $\hat H_0$ has delocalized quasiparticles, there is a third crossover or transition within the chaotic regime between the zero-dimensional quantum-dot-like regime~$(c)$ where the mean free path for interaction-induced scattering of the quasiparticles is longer than the system size, and a transport regime~$(d)$ where the mean free path is smaller than the system size so diffusive dynamics emerges.  (In cases with quenched randomness the transport can be subdiffusive~\cite{agarwal2017rare}, and in cases with long-range-in-real-space interactions it can be superdiffusive~\cite{joshi2021observing} (see also Ref.~\cite{PhysRevLett.127.057201}).)  Most conventional approaches to integrability-breaking, such as the Boltzmann equation~\cite{bastianello2021hydrodynamics}, describe regime~$(d)$. In regime~$(d)$, local thermalization is faster than transport across the system, and the latter process governs $\omega_{\mathrm{Th}}$. The behavior in $(c)$ followed by this crossover leads to a \emph{non-monotonic} dependence of $\omega_{\mathrm{Th}}$ on $N$ at fixed $\epsilon$ (Fig.~\ref{schematic}): the Thouless time first \emph{decreases} with system size as thermalization speeds up, before slowing down again as it becomes bottlenecked by transport. 

The slowest-relaxing mode, which sets the Thouless energy, has a qualitatively different character in regimes~$(c)$ and $(d)$. In regime $(c)$ it is a spatially uniform fluctuation of a charge that is conserved under $\hat H_0$; in regime $(d)$ it is a long-wavelength density fluctuation. Owing to the macroscopically different character of these two modes we conjecture that the crossover from $(c)$ to $(d)$ sharpens up in the dynamic limit considered here~\footnote{In the case where dynamics is diffusive, the longest-wavelength modes of the diffusion operator are plane-wave-like even when the system is microscopically disordered~\cite{PhysRevB.68.134207}.}.


Regime $(d)$ also includes the infinite system, where we can ask what processes dominate in setting the quasiparticle lifetime in the limit of small $\epsilon$.  For noninteracting $\hat H_0$, these can be low-order scattering processes, as in a standard Boltzmann equation scenario.  
For other systems, such as a XXZ spin chain perturbed in a way that ``breaks'' the Jordan-Wigner fermions (see Sec.~\ref{SecIV}), our numerical results are inconclusive, but exact diagonalization of small systems suggests that high-order processes dominate.


Note that for spatially extended systems where $\hat H_0$ is MBL and the interactions and hoppings in $\hat{H}$ are short-ranged enough in real space, the quantum-dot-like regime $(c)$ does not exist and the delocalization transition goes directly to the transport regime $(d)$.\cite{morningstar2021avalanches} 




\section{The Fock-space delocalization transition}
\label{SecII}
In this section we briefly review the analysis of AGKL for the delocalization transition in a quantum dot, then extend this analysis to the case of interacting integrable systems. 

\subsection{Review of AGKL analysis}

The AGKL paper considers the effects of interactions in a quantum dot with $N \gg 1$ fermionic levels, at generic intermediate filling $f$. Thus, $\hat H_0 = \sum_{i = 1}^N \varepsilon_i \hat c^\dagger_i \hat c_i$. Moreover, since the levels being analyzed are all within the single-particle Thouless energy of the dot, their eigenstates can be treated as random vectors. The many-body eigenstates of $\hat H_0$ are Fock states in which each such orbital is either occupied or unoccupied. These eigenstates can be represented as bit-strings, or equivalently as the vertices of an $N$-dimensional hypercube. The dot is perturbed by some generic, thermodynamically normalized, local four-fermion interaction $\epsilon\hat V = \frac{\epsilon}{N^{3/2}}\sum_{ijkl} V_{ijkl} \hat c^\dagger_i \hat c^\dagger_j \hat c_k \hat c_l$; since the single-particle eigenstates have no specific spatial structure, the matrix elements $V_{ijkl}$ can be treated as essentially random and order one. On the Fock-space hypercube, the interaction acts as a short-range hopping process, i.e., it connects each Fock state to $O(N^4)$ other states, related by moving any two electrons from filled orbitals to empty orbitals. 

For our purposes it is useful to generalize this discussion to the case where each application of $\hat V$ connects states that are separated by an order-one Hamming distance $n_h$ (so in the original AGKL analysis $n_h = 4$). For large $N$, the hopping graph is locally treelike, with coordination number $z \sim N^{n_h}$, effective hopping strength
\begin{equation}
\label{eq:ACAT}
t \sim  \frac{\epsilon}{N^{(n_h-1)/2}}~,
\end{equation}
and locally the Fock-space lattice is equivalent at large $N$ to a Bethe lattice.

By a classic result of Abou-Chacra, Anderson and Thouless\cite{ACAT}, the delocalization transition on a Bethe lattice with coordination number $z$ is known to occur at a critical hopping strength
\begin{equation}
t_c \sim \frac{w}{z \log{z}} \sim \frac{w}{N^{n_h}\log{N}},
\end{equation}
where $w$ denotes a typical energy change in $\hat H_0$ upon hopping, which we have set to order one. 

Combining the above expressions, we deduce that the critical perturbation strength $\epsilon=\epsilon_c(N)$ for Fock-space delocalization, and thus the critical perturbation strength required for thermalization scales with $N$ as
\begin{equation}
\label{eq:generalAGKL}
\epsilon_c(N) \sim \frac{1}{N^{(n_h+1)/2}\log{N}}.
\end{equation}
In Appendix \ref{Appendix}, we provide various perspectives on this result, including a self-contained derivation of the power-law contribution to this expression from estimating small denominators, together with a summary of AGKL's original derivation.



\subsection{Example: a perturbed all-to-all Ising model}

Before we turn to interacting integrable models with nontrivial quasiparticles, we introduce a version of the AGKL transition that is amenable to small-system numerical studies. For numerically accessible system sizes where one can fully diagonalize $H$, the AGKL model with $n_h = 4$ has the disadvantage that single hops are relatively ``long-range'' relative to the small Fock-space hypercube. To remedy this we consider a simpler model system that exhibits the phenomenology studied by AGKL at numerically accessible system sizes---namely an ensemble of random, perturbed all-to-all spin-$1/2$ Ising models given by
\begin{align}
\nonumber \hat H_{\mathrm{Ising}} &= \hat H_{0}+\epsilon \hat V, \quad \hat H_0 = \frac{1}{\sqrt{N}} \sum_{i\neq j}^N J_{ij} \hat \sigma_i^z \hat \sigma_j^z + \sum_{i=1}^N h_i \hat \sigma_i^z, \\
\label{eq:aaIsing}
\hat V &= \frac{1}{2}\frac{1}{\sqrt{N}} \sum_{i,j=1}^N h_{ij} \hat \sigma^z_i \hat \sigma^+_j + h.c.
\end{align}
with $J_{ij}$, $h_i$, $\mathrm{Re}[h_{ij}]$, $\mathrm{Im}[h_{ij}] \sim \mathcal{N}(0,1)$ identical and independently distributed (i.i.d.) normal variables. The eigenstates of the unperturbed Hamiltonian $\hat H_0$ are $\hat \sigma^z$ product states $|\sigma_1 \sigma_2 \ldots \sigma_N \rangle$ with $\sigma_i = \pm 1$.  These are the Fock states for this trivially ``integrable'' model.  The nearest-neighbors in Fock space are pairs of states that differ by only one spin flip.  We have added a perturbation $\hat V$ that breaks the ``integrability'' of $\hat H_0$ and produces ``hops'' in Fock space of distance only $n_h=1$, but we have used two-site terms to do this in $\hat V$, since $\hat H_0$ already had two-site interactions.  Note that the all-to-all interactions in both $\hat H_0$ and $\hat V$ are scaled by the appropriate power of $N$ to make them have a proper thermodynamic limit for $\epsilon$ of order one.  We will do this for all our models.  The crossover or transition to chaos (transition in this case) occurs in the {\it dynamic limit} at a coupling $\epsilon_c(N)$ that decreases to zero  in the large $N$ limit.  Thus all the models we will study are quantum chaotic when the {\it thermodynamic} limit is taken at any nonzero finite $\epsilon$.

For this model, the AGKL criterion~\eqref{eq:generalAGKL} gives:
%




\begin{equation}
\label{eq:isingscaling}
\epsilon_c(N) \sim \frac{1}{N \log{N}}.
\end{equation}

\begin{figure}[t]
    \centering
    \includegraphics[width=0.95\linewidth]{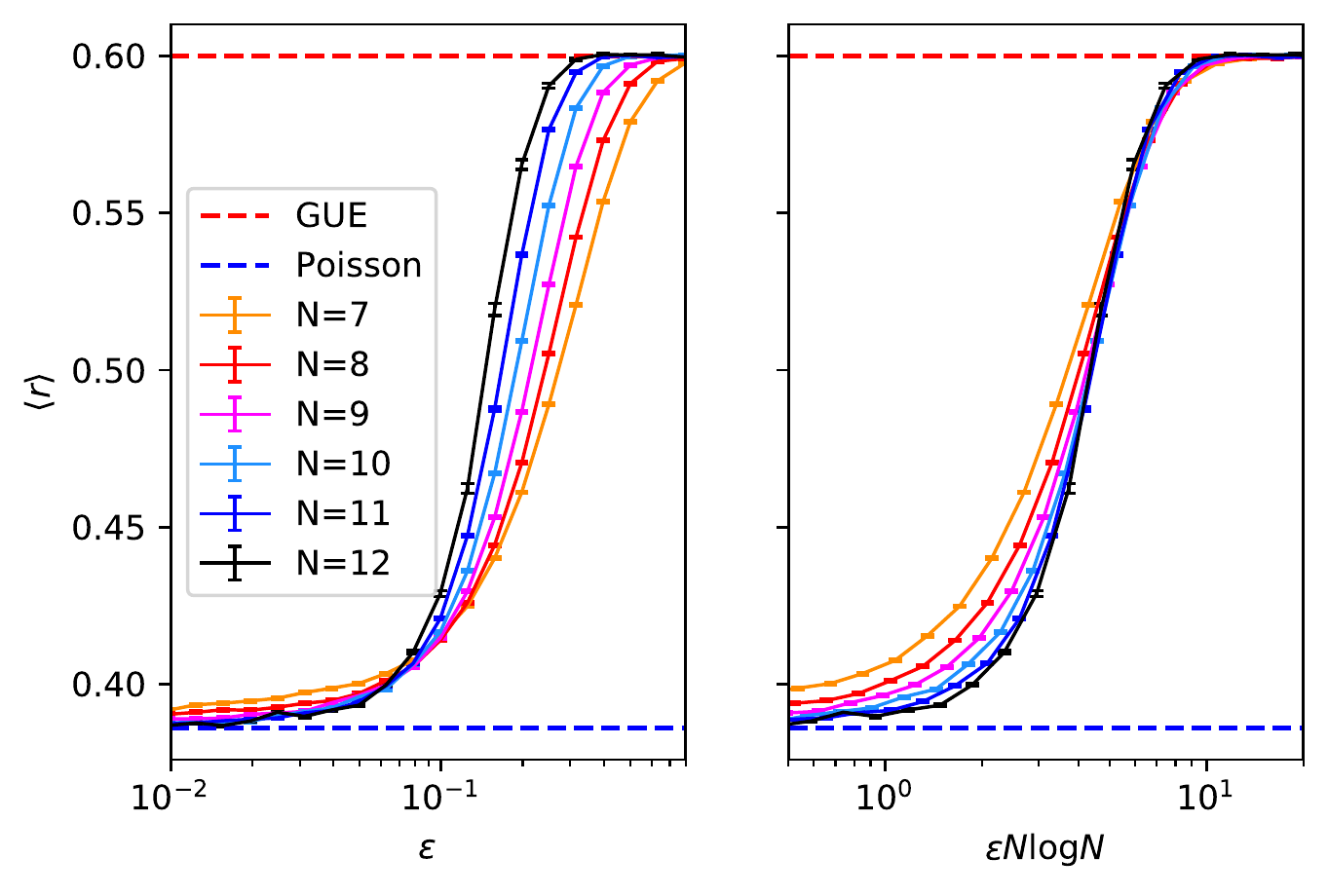}
    \caption{Finite-size behavior of $\langle r \rangle$ for the perturbed Ising model Eq. \eqref{eq:aaIsing}, scaled in the left panel according to the theoretical prediction Eq. \eqref{eq:isingscaling}. Each datapoint for $\langle r \rangle$ was obtained by simulating between $\mathcal{O}(100)$ and $\mathcal{O}(10000)$ realizations of the model Eq. \eqref{eq:aaIsing} for the largest and smallest system sizes respectively, and computing $\langle r \rangle$ from the middle $2/3$ of each spectrum. Since there are no degeneracies for generic couplings, we work in the full Hilbert space with dimension $d_N = 2^N$. Error bars denote the standard error of the sample mean of $\langle r \rangle$ over all realizations, which is typically $<0.1\%$ and at worst $<0.5\%$ for the largest systems simulated.}
    \label{Fig1}
\end{figure}

A particularly simple method for discerning the onset of chaos numerically is via level statistics. In the model Eq. \eqref{eq:aaIsing}, the ensembles of Hamiltonians corresponding to $\hat H_0$ and $\hat V$ are chosen to yield robust Poisson and GUE level statistics respectively, which are characteristic of generic integrable and (time-reversal-symmetry-breaking) chaotic quantum systems. Such level statistics are most directly observed using the $\langle r \rangle$ statistic\cite{HO}, which is defined as the average ratio (over the spectrum of some fixed $H$) of consecutive gaps
\begin{equation}
\langle r \rangle = \langle r_n \rangle, \quad r_n = \frac{\min{(\delta_n,\delta_{n+1})}}{\max{(\delta_n,\delta_{n+1})}},
\end{equation}
where $\delta_n = E_{n+1}-E_n$ denotes the separation between adjacent energy levels $\ldots>E_{n+1}>E_n>\ldots$ (these will generically be non-degenerate for the models and symmetry sectors considered in this paper). Scaling of $\langle r \rangle$ for the Hamiltonian Eq. \eqref{eq:aaIsing} according to the prediction Eq. \eqref{eq:isingscaling} is depicted in Fig. \ref{Fig1}. We see that this Hamiltonian interpolates between Poisson and GUE level statistics, for which $\langle r \rangle \approx 0.38$ and $\langle r \rangle \approx 0.6$ respectively, and appears to have a transition between these extremes that sharpens with increasing $N$ with respect to the scaling Eq. \eqref{eq:isingscaling}, and is thus consistent with a phase transition in this large-$N$ dynamic limit. One curious feature of Fig. \ref{Fig1} is that the curves collapse nearer the GUE value of $\langle r \rangle$ than the Poisson value of $\langle r \rangle$. We believe that this is a finite-size effect, and emphasize that it is the clear steepening of these curves with $N$ that provides the strongest numerical evidence for a phase transition as $N \to \infty$ in this model, among the models studied in this paper.

\begin{figure}[t]
    \centering
    \includegraphics[width=0.8\linewidth]{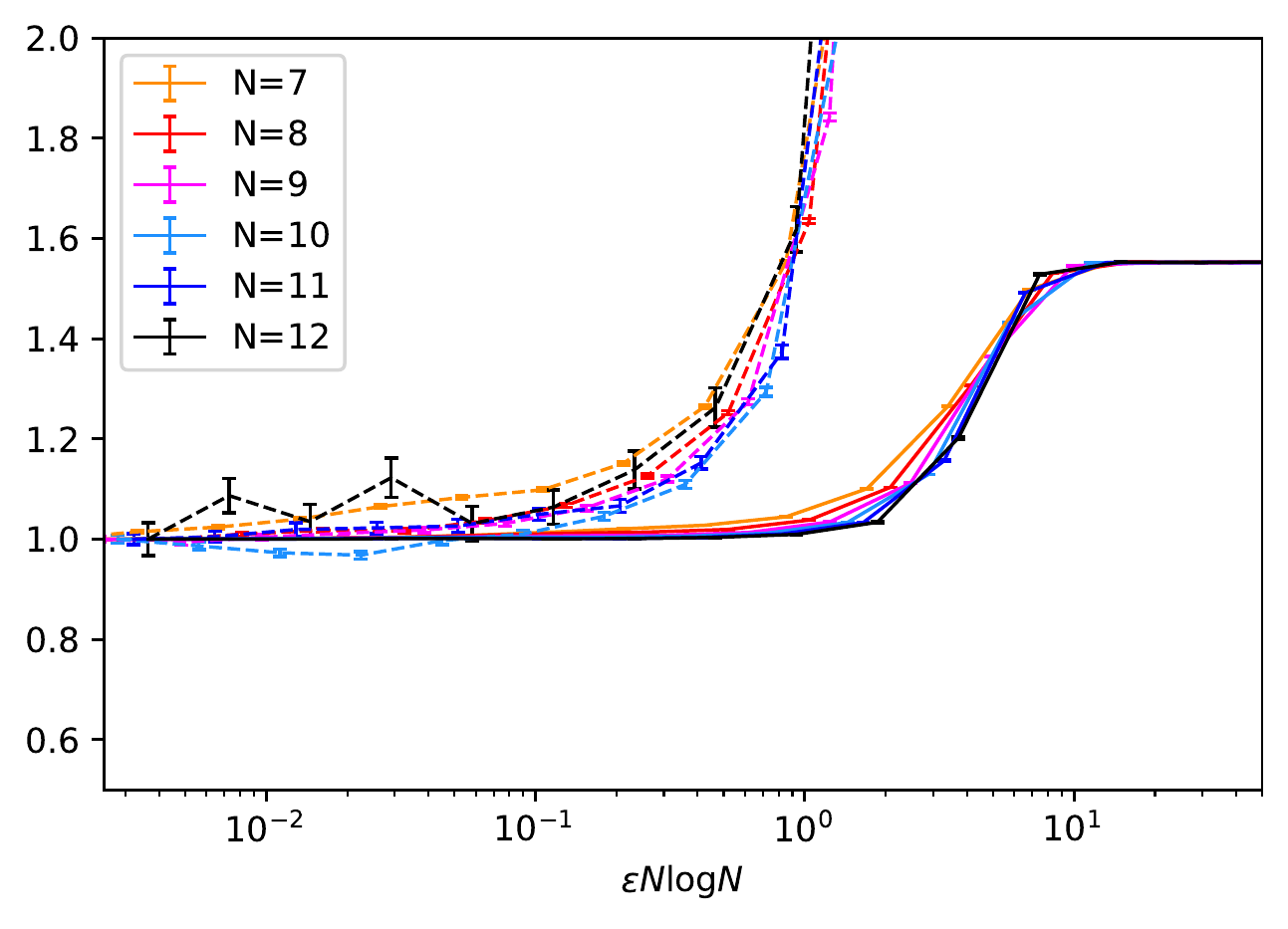}
    \caption{Finite-size behavior of $\langle r \rangle$ (solid lines) and $\mathrm{min}(r)$ (dashed lines) for the perturbed Ising model Eq. \eqref{eq:aaIsing}, scaled according to the theoretical prediction Eq. \eqref{eq:isingscaling}. Simulation details as in Fig. \ref{Fig1}. All curves shown are rescaled vertically by their values at the smallest $\epsilon = 2^{-13}$ simulated.}
    \label{Fig2}
\end{figure}

It is also of interest to consider the crossover between the dynamical regimes (a) and (b) of Fig. \ref{schematic} for the AGKL transition. This can be probed numerically by studying the minimum $\mathrm{min}(r) = \mathrm{min}_n r_n$ over each sample. As we argue in Appendix \ref{Appendix}, the Fock-space delocalization model of AGKL predicts that the onset of the very first resonances between adjacent energy levels is also set by the critical perturbation strength in Eq. \eqref{eq:ACAT}, but with a renormalized effective coordination number for Fock-space hopping, $\tilde{z} = 2^{2n_h}z$. Thus we expect that $\mathrm{min}(r)$ should exhibit 
the same scaling form
as $\langle r \rangle$, although with a very different scaling function, which is consistent with what is observed numerically in Fig. \ref{Fig2} (note that this observation is complicated by the fact that relative fluctuations in the minimum $\mathrm{min}(r)$ are substantially larger than those in $\langle r \rangle$ for the same number of samples). More generally, whenever an integrability-breaking perturbation induces purely local hopping in the Fock space of $\hat H_0$, we expect that diagnostics of quantum chaos that are sensitive to rare resonances, such as the adiabatic gauge potential\cite{PolkovnikovSels}, should begin to deviate from their $\epsilon = 0$ values at perturbation strengths that are polynomially small in the system size.

\subsection{General case: Fermi's Golden Rule}\label{fgrss}

We now turn to the case of general interacting integrable systems, solvable via the Bethe ansatz. Note that this class of systems includes both local spin chains and all-to-all models such as Gaudin magnets and central spin models~\cite{RevModPhys.76.643}. For these models, the coordinate Bethe ansatz provides a labeling system for eigenstates, in terms of the quantum numbers of occupied quasiparticles. These state labels can be expressed either in terms of half-integer Bethe quantum numbers (for local systems) or state-dependent rapidities~\cite{takahashi2005thermodynamics}; alternatively, one can label them by their eigenvalues under each of the local and quasi-local conserved charges~\cite{PhysRevLett.115.157201}. There are multiple different intuitively plausible ways to define a metric in this state space. However, while integrability-breaking perturbations once again induce matrix elements among the eigenstates, these matrix elements are no longer strictly bounded in Hamming distance under any of these obvious metrics. Although we expect that physical locality of $\hat V$ imposes some constraints on its matrix elements (for example, matrix elements between states with very different values of a local conserved quantity should be exponentially suppressed), the nature of these constraints is not obvious. 

Instead of working with an \emph{a priori} metric, we will reformulate the transition vs. crossover question as follows. A typical eigenstate of the integrable system is coupled to other eigenstates with a distribution of matrix elements $P(t)$. The typical matrix element must decay at least exponentially with system size, but there can be rare matrix elements that are much larger. We can ask whether first-order hybridization resonances become more or less likely as we consider increasingly small matrix elements (i.e., as we go to increasingly late times). For the onset of chaos to be a transition, the likeliest first-order resonances must involve atypically large matrix elements, which decay sub-exponentially with system size. One can heuristically regard this procedure as using the matrix element size as a proxy for Hamming distance, but we emphasize that it does not rely on any assumptions regarding the geometry of state space.


It is helpful to consider a toy model in which the probability distribution of matrix elements follows a power-law distribution, $P(t) \sim 1/t^{\gamma}$. (The power-law functional form is chosen for simplicity, although it appears to be consistent with our numerical results for a perturbed XX chain presented in Sec. \ref{SecIV}.) We now look for resonant transitions out of this eigenstate (i.e., transitions for which the matrix element exceeds the energy denominator), working our way from the highest to the lowest energies. At energy scale $\omega$, we can ignore matrix elements $t < \omega$, so the initial eigenstate is coupled to $N(\omega) \equiv \int_{\omega}^\infty P(t) dt \sim \omega^{1-\gamma}$ other states. The level spacing among these states scales as $\omega^{\gamma - 1}$; if $\gamma > 2$, this level spacing asymptotically becomes smaller than the matrix element $\omega$, and the initial state is unstable to long-range resonances at lowest order in perturbation theory. In this scenario, the relevant level spacing for delocalization is that of the slowest transitions, and this is set by the level spacing of the full system, i.e., it is exponential in $N$. By contrast, when $\gamma < 2$, the likeliest resonances involve the largest matrix elements; the nature of these depends on microscopic details but they will generically decay either algebraically or exponentially in $N$.

Note that there are two regimes within the crossover: $2 < \gamma < 3$ and $\gamma > 3$. In the former regime, the total spectral weight of $\hat V$, given by $\int t^2 P(t) dt$, is dominated by transitions with a large matrix element. In this regime, as one increases the perturbation strength $\epsilon$ past the threshold for chaos, the relaxation rate speeds up as ``faster'' thermalization channels, involving stronger transitions, become accessible. In this regime, at the level of the Golden Rule, the critical thermalization rate at the onset of chaos scales as $\Gamma_c \sim \epsilon^{(\gamma-1)/(\gamma-2)}$. On the other hand, when $\gamma > 3$, long-range transitions dominate the spectral sum, so the perturbation acts effectively like a random matrix, and the thermalization rate is only weakly affected by the opening of additional decay channels.


We now return to the question of probing the matrix element distribution numerically, with no assumptions about the underlying distribution of matrix elements. The idea behind this approach is to estimate when the broadening in the linewidth of Fock states due to dynamically introduced integrability-breaking perturbations becomes comparable to their level spacing. This provides an estimate of the perturbation strength at which a quasiparticle picture ceases to be meaningful. We note that similar considerations were used by AGKL to demarcate the Fock-space delocalization transition in quantum dots\cite{AGKL}.

Given a perturbed integrable system $H = \hat H_0 + \epsilon V$, we proceed as follows.
%
%
First let $|a\rangle$ denote an eigenstate chosen uniformly at random from the mid-spectrum of $\hat H_0$ (defined as e.g. the middle $2/3$ of its spectrum) and consider its matrix elements $V_{ba} = \langle b | V | a\rangle$ to other eigenstates in the mid-spectrum of $\hat H_0$. Ranking the squared matrix $|V_{ba}|^2$ in decreasing order for fixed $a$ and all mid-spectrum $b$ with $|E_a - E_b| < \Omega$ for some fixed $O(1)$ energy scale $\Omega$, averaging over all $a$ and then averaging over random couplings yields a ranked list of mean-squared off-diagonal matrix elements $|V_1|^2 > |V_2|^2 > \ldots > |V_{M}|^2$ in the mid-spectrum.

Each such matrix element $|V_m|^2$ defines a ``batch'' of other matrix elements $|V_n|^2 \sim |V_m|^2$ of the same order of magnitude. Concretely, we can define this batch as the set of $n$ such that $n \geq m$ and $|V_m|^2/e^2 < |V_n|^2 $. Letting $N_{\mathrm{batch}}(m)$ denote the number of $|V_n|^2$ in this batch, the decay rate to $m$ predicted by Fermi's golden rule is given by
\begin{align}
\nonumber \Gamma_m &= 2 \pi \epsilon^2 \frac{\left(\frac{1}{N_{\mathrm{batch}}(m)}\sum_{n \in \mathrm{batch}(m)} |V_n|^2\right)}{\Omega/N_{\mathrm{batch}}(m)} \\
&= \frac{2 \pi \epsilon^2}{\Omega} \sum_{n \in \mathrm{batch}(m)} |V_n|^2
\end{align}
since in the absence of level repulsion, the mean level spacing in this batch is expected to be $(\Delta E)_m \sim \Omega/N_{\mathrm{batch}}(m)$. We note that the condition for the validity of Fermi's Golden Rule in this setting is just the usual condition for the validity of Fermi's Golden Rule, namely that the batch of possible final states under consideration is not too small.

Comparing this decay rate to the mean level spacing yields the dimensionless perturbation strength
\begin{equation}
\frac{\Gamma_m}{(\Delta E)_m} = \frac{2 \pi \epsilon^2}{\Omega^2} N_{\mathrm{batch}}(m)\sum_{n \in \mathrm{batch}(m)} |V_n|^2.
\end{equation}
Maximizing this with respect to $m$ yields an estimate of the dominant decay channel due to $\hat V$; in particular, letting $m_*$ denote the argument of this maximum yields an estimate of the critical perturbation strength $\epsilon$, namely
\begin{equation}
\label{eq:FGR}
\epsilon^{\mathrm{FGR}}_c \sim \frac{\Omega}{\sqrt{2\pi}} \frac{1}{\sqrt{N_{\mathrm{batch}}(m_*)\sum_{n \in \mathrm{batch}(m_*)} |V_n|^2}}.
\end{equation}

The benefit of Eq. \eqref{eq:FGR} is that it provides an easily computable diagnostic of the threshold at which quasiparticles decay due to $\hat V$ and integrability is broken.


\subsection{Numerical diagnostics for Fock-space delocalization}

\label{sec:FGR}
The above discussion suggests two complementary numerical approaches for discerning a Fock-space delocalization transition in interacting integrable systems. The first is simply plotting $\langle r \rangle$ against $\epsilon$, as in Fig. \ref{Fig1}, and using this to deduce a ``critical'' $\epsilon^{\langle r \rangle}_{c}(N)$, either from the crossing point of the curves for $\langle r \rangle$ if this is visible, or by numerically extracting $\epsilon^{\langle r \rangle}_{c}(N)$ at which $\langle r \rangle \approx 0.49$ (halfway between Poisson and GUE) if there is no clear crossing point. In this paper we will always consider GUE perturbations, with a view to enhancing the change in $\langle r \rangle$ (there is no reason to expect any of our conclusions to be altered for GOE perturbations, beyond their smaller value of $\langle r \rangle \approx 0.53$). This diagnostic refers solely to the level statistics of the Hamiltonian in question. 

The second approach is plotting the Fermi golden rule estimate Eq. \eqref{eq:FGR} against system size $N$, and checking for polynomial versus exponential decay in $N$. This diagnostic refers to matrix elements of the perturbation $\hat V$ with respect to eigenfunctions of the unperturbed Hamiltonian $\hat H_0$, and in this sense is complementary to the level statistics analysis.

When both estimates $\epsilon^{\langle r \rangle}_c$ and $\epsilon^{\mathrm{FGR}}_c$ decay polynomially in $N$, this is evidence that integrability-breaking for the model in question is driven by a Fock-space delocalization phase transition in the appropriate dynamic limit, with the power-law in Eq. \eqref{eq:generalAGKL} indicating the effective Fock-space hopping distance $n_h$.  When both estimates decay exponentially in $N$, this is evidence that integrability-breaking exhibits a crossover rather than a transition and occurs via a Fock-space hopping distance $n_h \sim N$.  This is because exponential decay of the Fermi golden rule threshold Eq. \eqref{eq:FGR} in $N$ implies that $\hat V$ mixes each many-quasiparticle state with exponentially many other states in the many-body Hilbert space, ruling out a localized phase as $N\to \infty$ for any sub-exponential rescaling of $\epsilon$ with $N$.

\section{Perturbed integrable quantum dots}
\label{SecIII}
\subsection{Interacting integrable quantum dots}
Let us now apply the above considerations to integrability-breaking perturbations of interacting, integrable quantum dots. Note that this generalizes the analysis of AGKL\cite{AGKL}, who considered such perturbations for non-interacting quantum dots. The difference between a non-interacting and an interacting integrable quantum dot is captured by the quantum numbers, or ``rapidities'' defining the state. In a non-interacting quantum dot, each state is uniquely determined by the occupation numbers of a fixed set of rapidities. In an interacting integrable quantum dot, these rapidities depend on the state; their mutual interactions are captured by the Bethe equations. 

To be concrete, let us take as our interacting, integrable quantum dot the following transverse spin-$1/2$ Gaudin model 
\begin{equation}
\label{eq:pureGaudin}
\hat H_{\mathrm{Gaudin}} = \frac{1}{\sqrt{N}} \sum_{i \neq j} J_i^* J_j \hat S_i^{+} \hat S_j^{-}.
\end{equation}
This model is integrable for all choices of the couplings $J_i\in\mathbb{C}$. It has a global $U(1)$ symmetry and a ``hidden'' time-reversal symmetry, achieved by rotating each spin independently about the $z$-axis to cancel the phase of its complex coupling $J_i$. 

Local integrable systems, such as the spin-$1/2$ XXZ chain, are characterized by an extensive number of local conserved charges. In such systems, the Hamiltonian $\hat H = \sum_{i=1}^N \hat h_i$ can be written as a sum over local terms $h_i$, and there exist extensively many other linearly independent operators $\hat Q^{(n)} = \sum_i \hat q^{(n)}_i,\, n=2,\ldots,N$ with $\hat q^{(n)}_i$ local that are conserved and mutually commuting, i.e.
\begin{equation}
[\hat Q^{(m)}, \hat Q^{(n)}] = 0, \quad [\hat Q^{(m)}, \hat H] = 0, \quad m,n=2,\ldots, N.
\end{equation}

For all-to-all models such as quantum dots, the notion of spatial locality is no longer natural. Instead, integrability in such systems is characterized by the existence of extensively many conserved charges that are ``2-local'' in the complexity theory sense. Gaudin's construction of these models proceeds by solving for sets of commuting spin bilinears with a specific structure; for the model Eq. \eqref{eq:pureGaudin}, these take the form
\begin{equation}
\hat G^{(i)} = \sum_{k \neq i} \frac{1}{2}\left(v_{ik} \hat S_i^- \hat S_k^+ + v_{ik}^* \hat S_i^+ \hat S_k^-\right) + w_{ik}\hat S_i^z \hat S_k^z,
\end{equation}
with the coefficients $v$, $w$ chosen to satisfy the condition $[\hat G^{(i)},\hat G^{(j)}]=0$. Assuming that $v^*_{ji} =-v_{ij}$, these conditions imply the ``Gaudin equations'',
\begin{equation}
w_{ik} - w_{ij} = \frac{v_{ik}v_{ji}}{v_{jk}}
\end{equation}
for pairwise distinct $i$, $j$ and $k$, which are solved by
\begin{equation}
v_{ij} = \frac{J_i J_j^*}{|J_i|^2-|J_j|^2}, \quad w_{ij} = \frac{|J_j|^2}{|J_i|^2 - |J_j|^2}
\end{equation}
for any distinct $J_i \in \mathbb{C}$. In terms of these operators, a family of all-to-all Hamiltonians with an extensive number of mutually commuting, 2-local conserved charges is given by
\begin{equation}
\hat H = \sum_{i=1}^N a^{(1)}_i \hat G^{(i)}, \quad \hat Q^{(n)} = \sum_{i=1}^N a^{(n)}_i \hat G^{(n)},
\end{equation}
where $a^{(n)}_i$ are the coefficients of any invertible real matrix. The choice $a^{(1)}_i = 2|J_i|^2/N^{1/2}$ recovers the Hamiltonian Eq. \eqref{eq:pureGaudin} that we will study in this section.

Integrable spin models of this type were introduced by Gaudin\cite{Gaudin}, building on earlier work by Richardson\cite{Richardson2}. The model Eq. \eqref{eq:pureGaudin} lies in a more general class than the models originally considered by Gaudin; this class of ``non-skew'' Gaudin models first arose in the context of nuclear physics\cite{FirstNonSkew}, with several subsequent studies\cite{Skrypnyk2005,Balantekin2005,Lukyanenko,WishartSYK,Claeys2019}. Recently, an experimental protocol for realizing Eq. \eqref{eq:pureGaudin} (with real couplings $J_i \in \mathbb{R}$) in systems of atoms coupled to an optical cavity was proposed\cite{EhudGaudin}.

\subsection{Quasiparticle Fock space for the Gaudin model}
To make contact with the AGKL intuition for non-interacting quantum dots, it is helpful to express integrability of the Gaudin model in terms of ``quasiparticle operators'', which are analogous to pair creation and annihilation operators in the Richardson model\cite{Richardson2}. This formalism also leads directly to the model's exact Bethe-ansatz solution.

Our starting point is the so-called ``rational Gaudin algebra'' associated with the Hamiltonian Eq. \eqref{eq:pureGaudin}, which is determined by the couplings $J_i$ and a continuous parameter $z$, such that\cite{Gaudin,FirstNonSkew,Balantekin2005}
\begin{align}
\hat B^{+}(z) = \sum_{i=1}^N \frac{J_i^*}{1-|J_i|^2 z} \hat S_i^{+},\\
\hat B^{-}(z) = \sum_{i=1}^N \frac{J_i}{1-|J_i|^2 z} \hat S_i^{-},\\
\hat B^z(z) =  \sum_{i=1}^N \frac{|J_i|^2}{1-|J_i|^2 z} \hat S_i^z.
\end{align}
These operators satisfy the commutation relations
\begin{align}
\label{CR}
[\hat B^+(z), \hat B^-(0)] &= 2 \hat B^z(z), \\
[\hat B^z(z), \hat B^{\pm}(w)] &= \pm \frac{\hat B^{\pm}(z) - \hat B^{\pm}(w)}{z-w}.
\end{align}
In terms of these operators, the Hamiltonian Eq. \eqref{eq:pureGaudin} is given by
\begin{equation}
\hat H_{\mathrm{Gaudin}} = \frac{1}{\sqrt{N}} \left[\hat B^+(0) \hat B^-(0) - \hat B^z(0) - \frac{1}{2}\sum_{i=1}^N |J_i|^2\right].
\end{equation}
From the commutation relations Eqs. \eqref{CR}, one can construct the exact $M$-particle eigenstates
\begin{equation}
|z_1 z_2\ldots z_M \rangle = \hat B^+(z_1) \hat B^+(z_2) \ldots \hat B^+(z_M) | 0 \rangle
\end{equation}
by acting repeatedly on the pseudovacuum $|0\rangle$ with all spins down, with the ``rapidities'' $z_a$ required to satisfy the $M$ coupled Bethe equations
\begin{equation}
\label{eq:BetheEq}
\sum_{b \neq a}^M \frac{2}{z_a-z_b} + \sum_{j=1}^N \frac{|J_j|^2}{1-|J_j|^2 z_a} + \frac{1}{z_a} = 0, \quad a=1,2,\ldots,M,
\end{equation}
and the energy given by
\begin{equation}
E(z_1,z_2,\ldots,z_M) = -\frac{1}{\sqrt{N}} \sum_{a=1}^M \frac{1}{z_a}.
\end{equation}
This formalism allows for a precise definition of a ``quasiparticle Fock space'', as the set of vectors $\vec{z} \in \mathbb{C}^M$ satisfying the Bethe equations Eq. \eqref{eq:BetheEq} for each $M = 0,1,\ldots,N$, up to rearrangements of the $z_i$. The validity of the AGKL Fock-space delocalization model for few-site perturbations of this system hinges on how far single-site operators $\hat S_i^{\pm}$ 
induce local hopping in this quasiparticle Fock space. Based on the fact that the single-site operators $\hat S_i^{\pm}$ are simply linear combinations of the $\hat B^{\pm}(z)$, one might conjecture that they generate hops at a Hamming distance $n_h=1$ in this Fock space, and therefore induce a Fock-space delocalization transition.
Below we present numerical evidence in favour of this claim for accessible system sizes.

\subsection{Few-spin perturbations of the transverse Gaudin model}

\begin{figure}[t]
    \centering
    \includegraphics[width=0.99\linewidth]{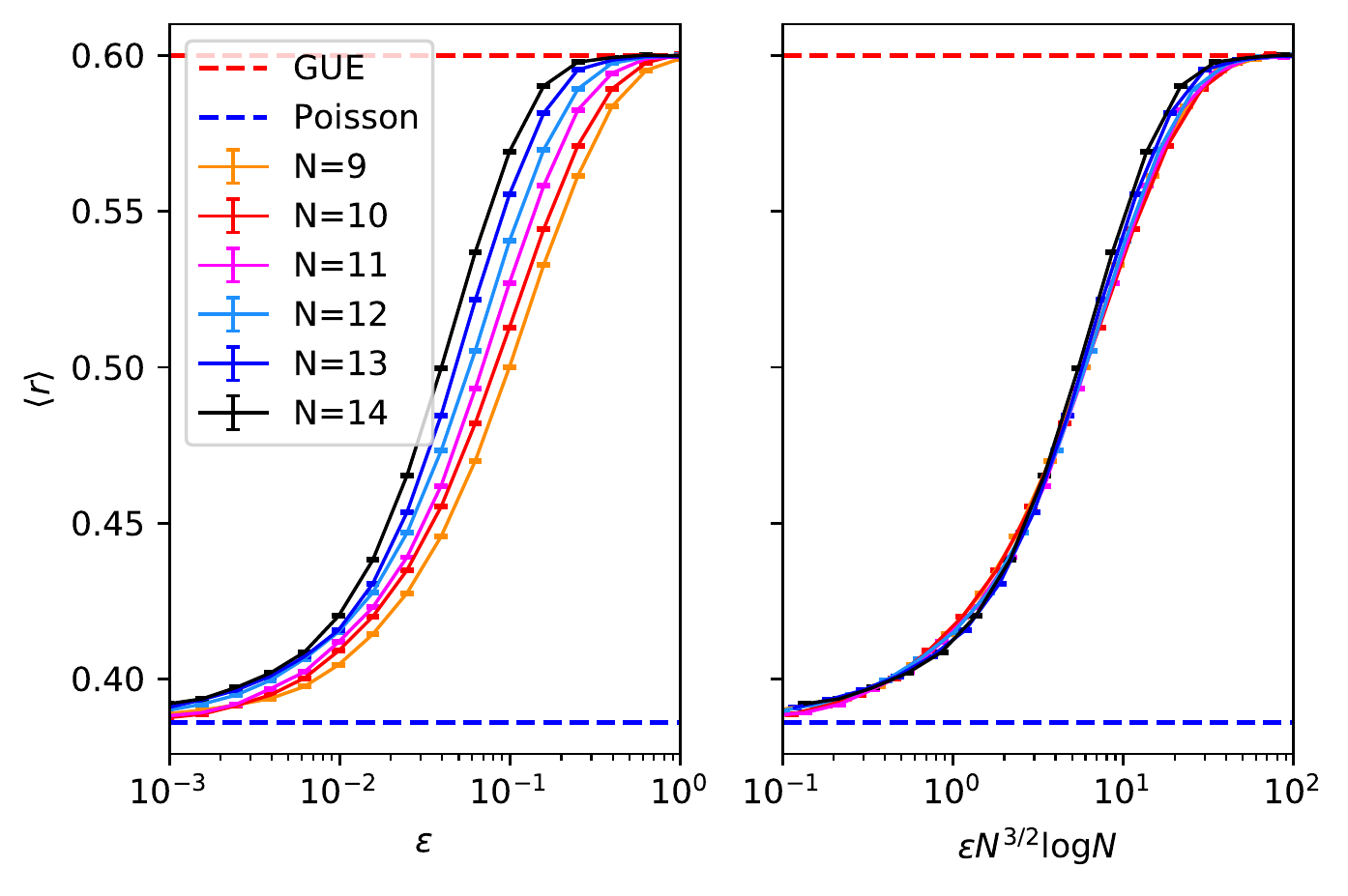}
    \caption{Scaling collapse of $\langle r \rangle$ for the transverse Gaudin model perturbed by a two-spin perturbation Eq. \eqref{eq:Gaudin2}, according to the theoretical prediction Eq. \eqref{eq:Gaudin2scaling}. Each datapoint for $\langle r \rangle$ was obtained by simulating between $\mathcal{O}(100)$ and $\mathcal{O}(10000)$ realizations of the model Eq. \eqref{eq:Gaudin2} for the largest and smallest system sizes respectively, and computing $\langle r \rangle$ from the middle $2/3$ of each spectrum. Since the model is $U(1)$ symmetric, we work in the approximately half-filled sector with $S^+$ occupation number $\lfloor N/2\rfloor$ and Hilbert space dimension $d_N = \begin{pmatrix} N \\ \lfloor N/2 \rfloor\end{pmatrix}$. Error bars denote the standard error of the sample mean over all realizations, which is $<0.1\%$ for all data points shown.}
    \label{Fig3}
\end{figure}

We now consider various random few-site perturbations of the transverse Gaudin model $\hat H_{\mathrm{Gaudin}}$, first the random two-site perturbation
\begin{align}
\label{eq:Gaudin2}
\nonumber &\hat H_{\mathrm{Gaudin}+2} = \hat H_0 + \epsilon \hat V, \quad \hat H_{0} = \frac{1}{\sqrt{N}} \sum_{i \neq j} J_i^* J_j \hat S_i^{+} \hat S_j^{-},\\
&\hat V = 
\frac{1}{\sqrt{N}} \sum_{i<j}(J_{ij} \hat S_i^{+} \hat S_j^{-} + J_{ij}^* \hat S_j^{+} \hat S_i^{-}) + \sum_{j=1}^N h_j \hat S_j^z
\end{align}
with i.i.d. couplings $\mathrm{Re}(J_j)$, $\mathrm{Im}(J_j)$, $\mathrm{Re}(J_{ij})$, $\mathrm{Im}(J_{ij})$, $h_j \sim \mathcal{N}(0,1)$. Here the random two-site terms drive the model to GUE level statistics, while the random single-site $z$-field serves to remove an accidental time-reversal symmetry for even $N$. The two-site term is the dominant source of Fock-space hopping, and has $n_h = 2$ in our previous notation, so that the Fock-space delocalization model Eq. \eqref{eq:generalAGKL} predicts a critical perturbation strength
\begin{equation}
\label{eq:Gaudin2scaling}
\epsilon_c(N) \sim \frac{1}{N^{3/2}\log{N}}
\end{equation}
for the onset of chaos. As shown in Fig. \ref{Fig3} the collapse of $\langle r \rangle$ to this scaling law is excellent, and shows a slight steepening about its crossing point at the largest system sizes, indicative of a possible phase transition as $N \to \infty$.

\begin{figure}[t]
    \centering
    \includegraphics[width=0.99\linewidth]{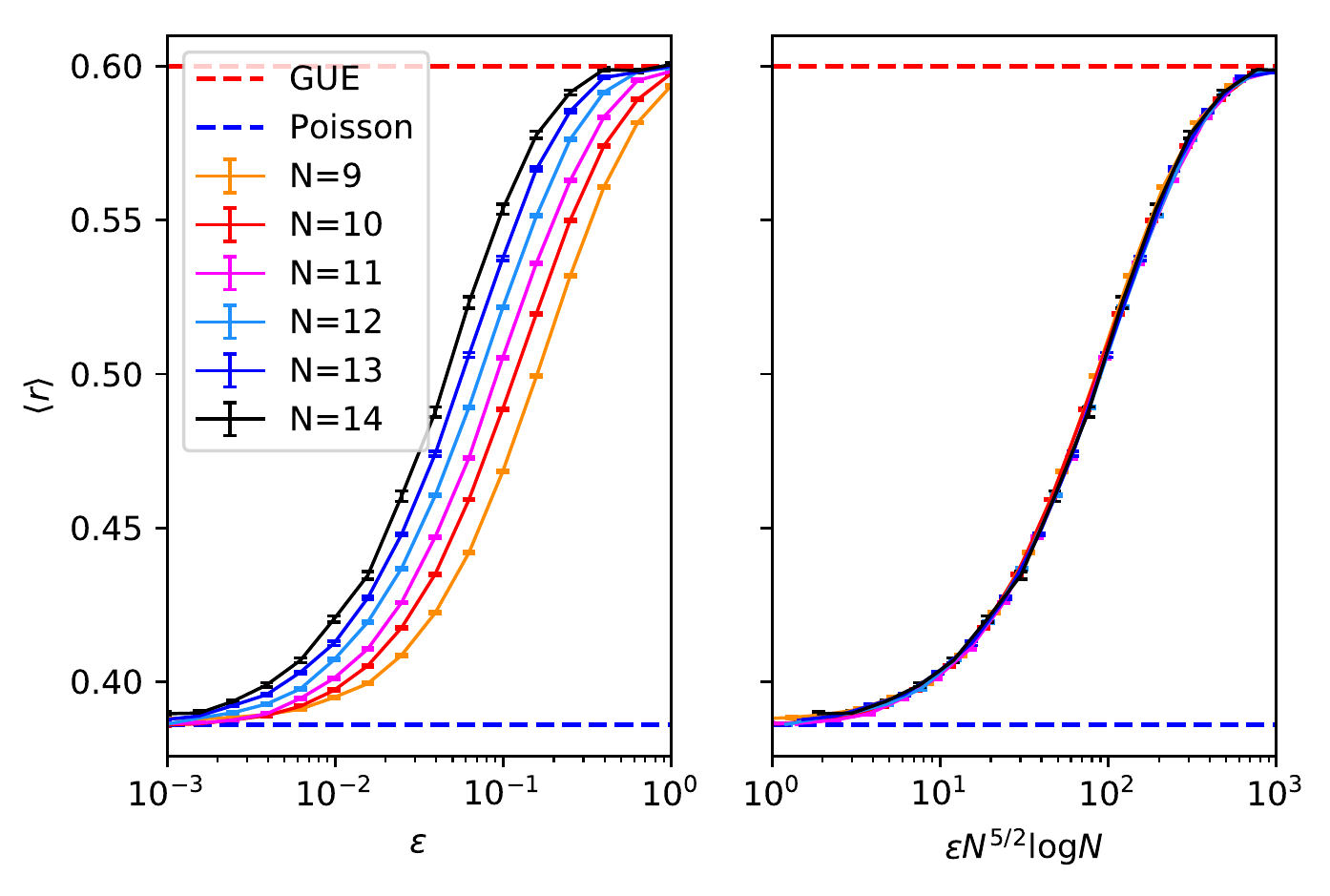}
    \caption{Scaling collapse of $\langle r \rangle$ for the transverse Gaudin model perturbed by a four-spin perturbation Eq. \eqref{eq:Gaudin4}, according to the theoretical prediction Eq. \eqref{eq:Gaudin4scaling}. Simulation details as for Fig. \ref{Fig3}. Error bars denote the standard error of the sample mean over all realizations, which is typically $<0.1\%$ and at worst $<0.4\%$ for the largest systems simulated.}
    \label{Fig4}
\end{figure}

We next consider the random four-site perturbation
\begin{align}
\label{eq:Gaudin4}
\nonumber &\hat H_{\mathrm{Gaudin}+4} = \hat H_0 + \epsilon \hat V, \quad \hat H_{0} = \frac{1}{\sqrt{N}} \sum_{i \neq j} J_i^* J_j \hat S_i^{+} \hat S_j^{-}, \\
&\hat V = \frac{1}{N^{3/2}} \sum_{i<j,k<l} J_{ijkl}\hat S_i^+ \hat S_j^+ \hat S^-_k \hat S^-_l + J_{ijkl}^* \hat S_l^+ \hat S_k^+ \hat S_j^- \hat S_i^- 
\end{align}
with i.i.d. couplings $\mathrm{Re}(J_j)$, $\mathrm{Im}(J_j)$, $\mathrm{Re}(J_{ijkl})$, $\mathrm{Im}(J_{ijkl}) \sim \mathcal{N}(0,1)$. Here, $n_h = 4$ and so Eq. \eqref{eq:generalAGKL} predicts a critical perturbation strength
\begin{equation}
\label{eq:Gaudin4scaling}
\epsilon_c(N) \sim \frac{1}{N^{5/2}\log{N}}
\end{equation}
for the onset of chaos. As shown in Fig. \ref{Fig4}, the $\langle r \rangle$ statistic exhibits an excellent collapse to this scaling law on accessible system sizes. However, there is not much evidence for steepening with $N$, presumably because the hopping distance $n_h=4$ in this case is only a few times smaller than the largest accessible system size, $N=14$. 

\begin{figure}[t]
    \centering
    \includegraphics[width=0.99\linewidth]{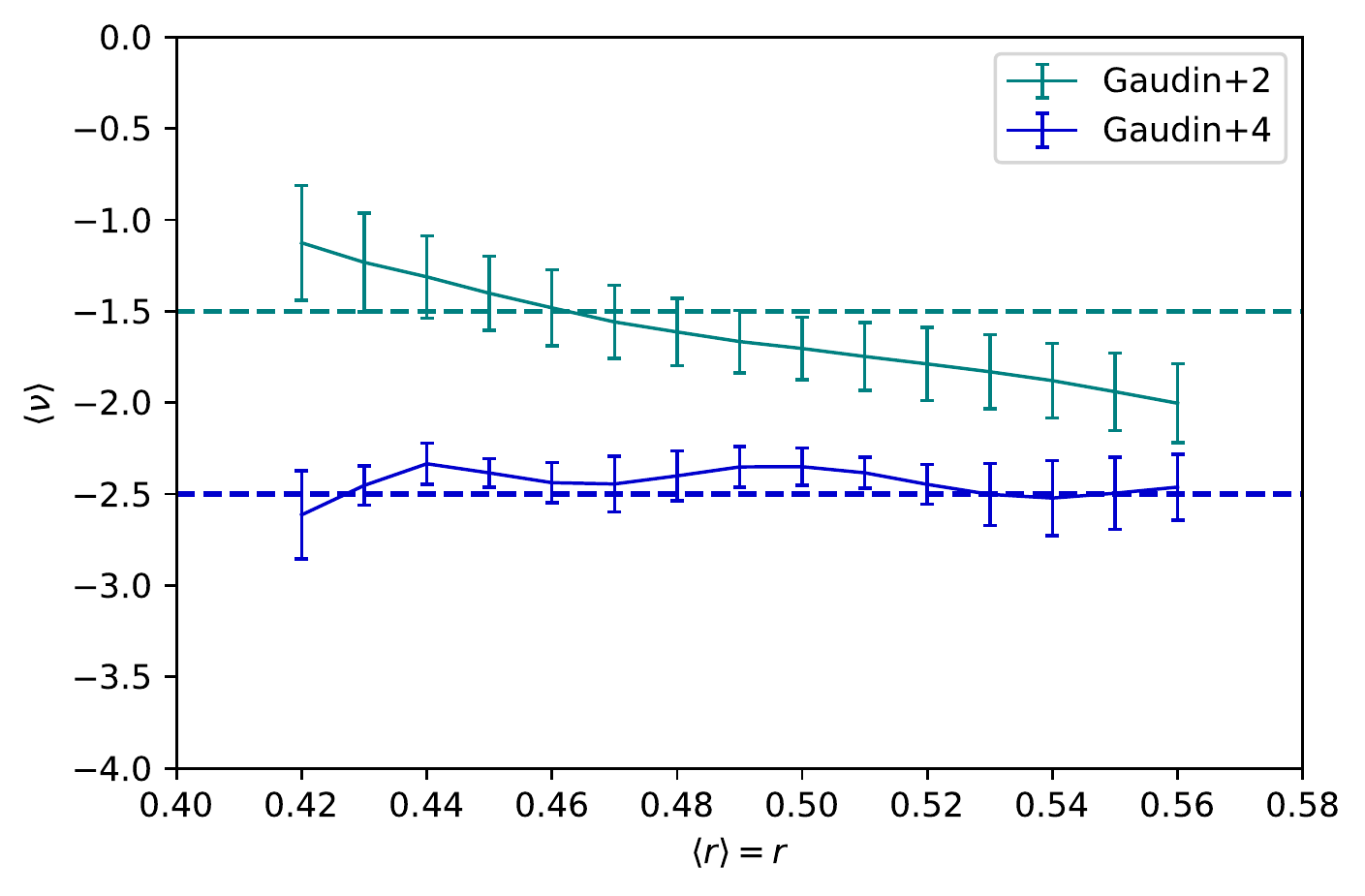}
    \caption{Power-law dependence of $\epsilon(r,N)\log{N}$ on $N$, as extracted from the numerical $\langle r\rangle$ data for the Hamiltonians $\hat H_{\mathrm{Gaudin}+2}$ and $\hat H_{\mathrm{Gaudin}+4}$, using the formula in Eq. \eqref{eq:extrap}. Dotted lines show the respective analytical predictions, Eqs. \eqref{eq:Gaudin2scaling} and \eqref{eq:Gaudin4scaling} for the exponent $\nu$ at the critical point. Error bars denote standard error of the sample mean Eq. \eqref{eq:extrap}, with $n=6$.}
    \label{Fig5}
\end{figure}

It is worth emphasizing that the power laws in Eqs. \eqref{eq:Gaudin2scaling} and \eqref{eq:Gaudin4scaling} can be deduced directly from the $\langle r \rangle$ data itself, and are clearly distinguishable within the sample error of a fit to the Fock-space delocalization ansatz Eq. \eqref{eq:generalAGKL} over accessible $N$. Concretely, we can study the $N$ dependence of the onset of chaos by defining the threshold $\epsilon(r,N)$ at which $\langle r \rangle$ takes on an intermediate value $\langle r \rangle =r \in (0.41,0.57)$ for each $N$. This is interpolated to unsampled values of $r$ by using an accurate spline (error within linewidth) of the raw data. From $\epsilon(r,N)$ we can estimate the exponent $\nu$ such that
\begin{equation}
\epsilon(r,N) \sim \frac{1}{N^\nu \log{N}}
\end{equation}
by averaging the gradient of $\log{(\epsilon(r,N) \log{N})}$ with respect to $\log{N}$, i.e. for $n$ consecutive system sizes $N=N_1,N_1+1,\ldots,N_n$ we define
\begin{equation}
\label{eq:extrap}
\langle \nu \rangle = \sum_{i=1}^{n-1} \frac{\log{[(\epsilon(r,N_{i+1})\log{N_{i+1}})/(\epsilon(r,N_i)\log{N_i}})]}{(n-1)\log {N_{i+1}/N_i}}.
\end{equation}
The resulting estimate for $\langle \nu \rangle$, together with the standard error of this average (note that $n=6$ for our simulations) is plotted in Fig. \ref{Fig5} for the perturbed interacting integrable models Eq. \eqref{eq:Gaudin2} and Eq. \eqref{eq:Gaudin4}. The first model seems to exhibit a steepening transition on either side of the theoretically predicted exponent from Fock-space delocalization Eq. \eqref{eq:generalAGKL}. For the model with the larger hopping range $H_{\mathrm{Gaudin+4}}$, no such steepening is visible for the reason discusssed above; correspondingly, the agreement with the theoretical prediction Eq. \eqref{eq:Gaudin4scaling} is particularly good.

\begin{figure}[t]
    \centering
    \includegraphics[width=0.99\linewidth]{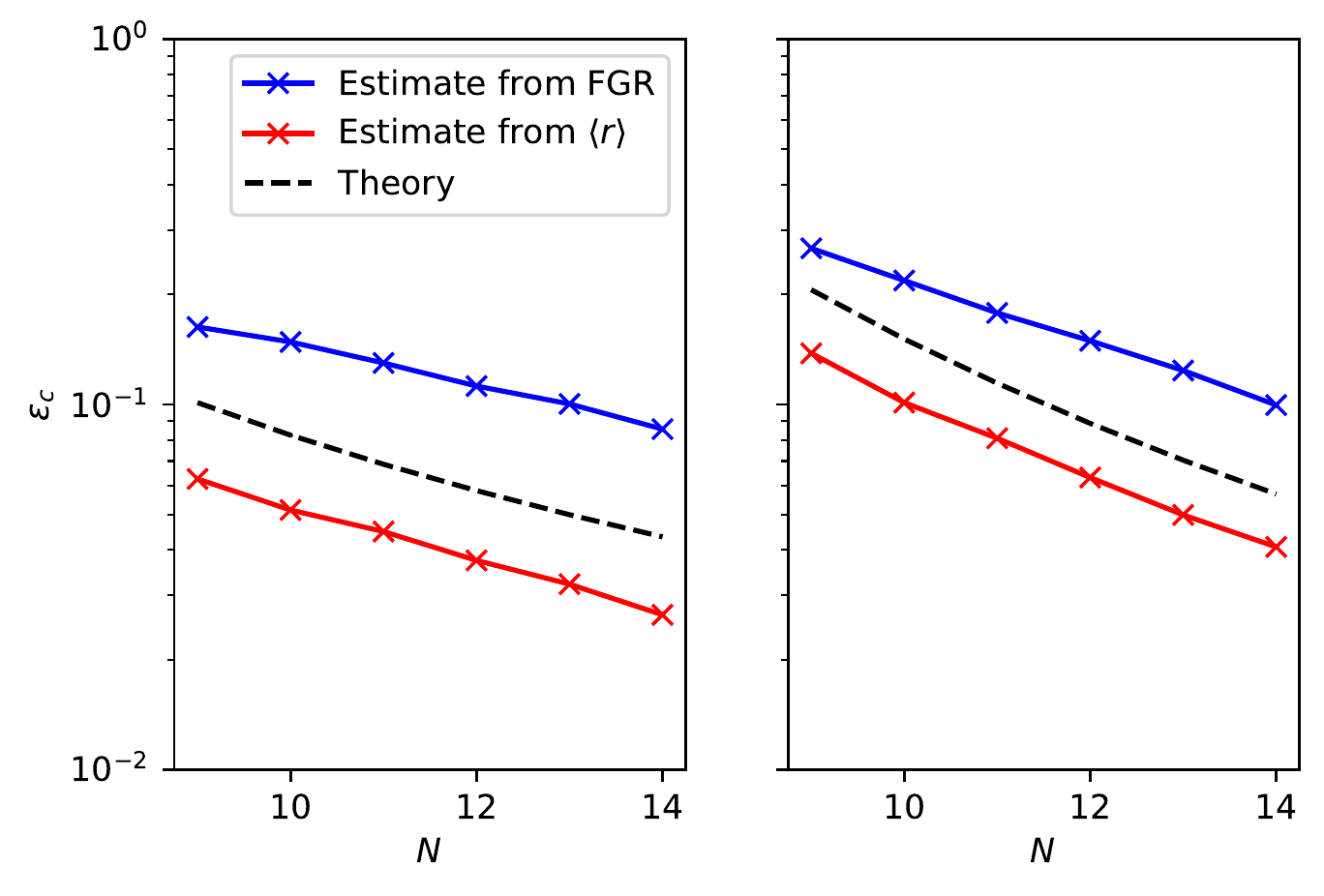}
    \caption{Thresholds for the critical perturbation strength extracted from a Fermi golden rule estimate Eq. \eqref{eq:FGR} and the numerical $\langle r \rangle$ data respectively, according to the prescription described in Section \ref{sec:FGR}, for the models Eqs. \eqref{eq:Gaudin2} (\textit{left}) and \eqref{eq:Gaudin4} (\textit{right}) respectively. Black dotted lines are guides to the eye, and directly proportional to the Fock-space delocalization predictions Eqs. \eqref{eq:Gaudin2scaling} and \eqref{eq:Gaudin4scaling}. For both models, both numerical estimates for $\epsilon_c$ track this theoretical prediction for accessible $N$.}
    \label{Fig6}
\end{figure}

This concludes our numerical demonstration that the level statistics of perturbed interacting quantum dots scale as expected from the arguments of AGKL. We next turn to the Fermi golden rule estimate, based on the decay rate of quasiparticles, that was discussed above. In particular, one can directly compare $\epsilon^{\langle r \rangle}_c$ and $\epsilon^{\mathrm{FGR}}_c$ for the interacting models, as discussed in Section \ref{sec:FGR}, and check for polynomial versus exponential decay with $N$. This check is depicted in Fig. \ref{Fig6}. We see that accessible system sizes are consistent with the polynomial decay predicted by a Fock-space delocalization model. Let us however emphasize that from numerics alone, it is impossible to rule out a very weak exponential decay that could eventually dominate the AGKL prediction at larger $N$, resulting in a crossover to quantum chaos rather than a transition.

\begin{figure}[t]
    \centering
    \includegraphics[width=0.95\linewidth]{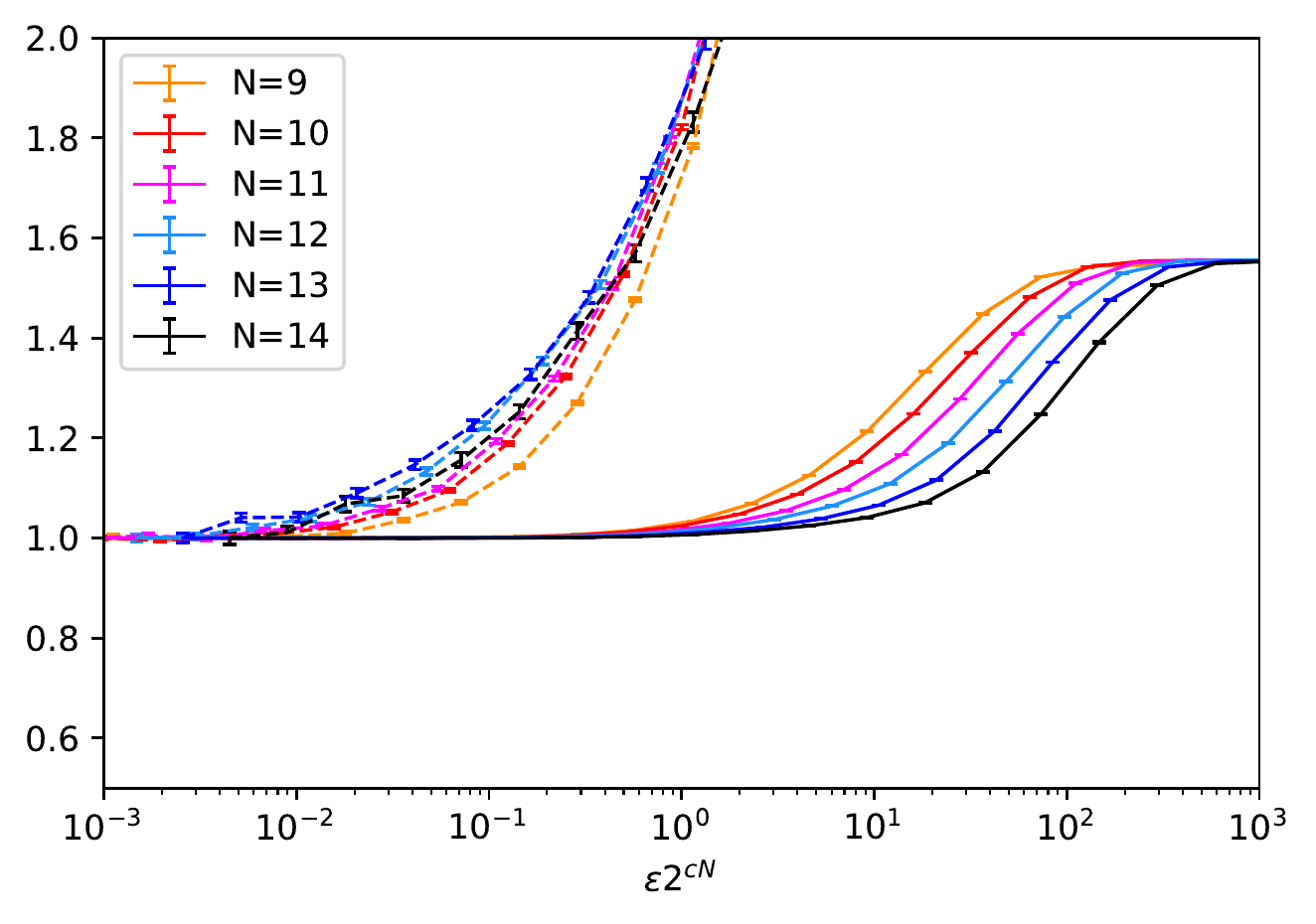}
    \caption{Finite-size behavior of $\langle r \rangle$ (solid lines) versus $\mathrm{min}(r)$ (dashed lines) for the Hamiltonian $\hat H_{\mathrm{Gaudin}+2}$, with the perturbation strength rescaled by an exponential $2^{cN}$ with $c=0.8$. Simulation details as in Fig. \ref{Fig3}. All curves are rescaled vertically by their value at the smallest perturbation strength $\epsilon = 2^{-19}$ simulated. It is clear that the exponential rescaling that induces a collapse of $\mathrm{min}(r)$ fails to collapse $\langle r \rangle$.}
    \label{Fig7}
\end{figure}

In both these models, the minimum $r$ behaves markedly differently from the mean $r$, seeming to exhibit regime change at a coupling strength that is exponentially small in the system size. This behaviour is consistent with earlier results on the adiabatic gauge potential in perturbed interacting integrable systems\cite{PolkovnikovSels}. However, as shown in Fig. \ref{Fig7} for $\hat H_{\mathrm{Gaudin}+2}$, such exponential scaling manifestly fails to collapse the behaviour of $\langle r \rangle$ between the extremes of integrability and chaos $0.38 < \langle r \rangle < 0.6$. The power law scaling of $\langle r \rangle$ depicted in Figs. \ref{Fig3} and \ref{Fig4} thus indicates a robust intermediate regime between the onset of many-body resonances and the onset of chaos, that is large relative to the conjectured perturbation strength $\epsilon_c(N) \sim 1/(N^\nu\log{N})$ for the onset of chaos in these models. One intuitive explanation for this behaviour, compared to the all-to-all Ising model discussed above, is that for perturbed Gaudin dots, the matrix elements of few-spin perturbations $\hat V$ have long tails in Fock space that allow some many-body resonances to form at exponentially small $\epsilon$, but sufficiently little weight in these tails that a delocalization transition in $\langle r \rangle$ persists at larger values of $\epsilon$.

Finally, we note that one can also break $U(1)$ symmetry of $\hat H_0$ while preserving integrability, by adding an onsite XY field of the form\cite{Lukyanenko}
\begin{equation}
\hat H_{0} = \frac{1}{\sqrt{N}} \sum_{i \neq j} J_i^* J_j \hat S_i^{+} \hat S_j^{-} + h_{XY} \sum_{i} (J_i \hat S_i^{-} + J^*_i \hat S_i^+),
\end{equation}
with $h_{XY}$ arbitrary. We have checked that this does not alter the observation of power-law scaling in $\langle r \rangle$ for few-site chaotic perturbations and accessible system sizes.

\subsection{Many-spin perturbations of the transverse Gaudin model}

\begin{figure}[t]
    \centering
    \includegraphics[width=0.99\linewidth]{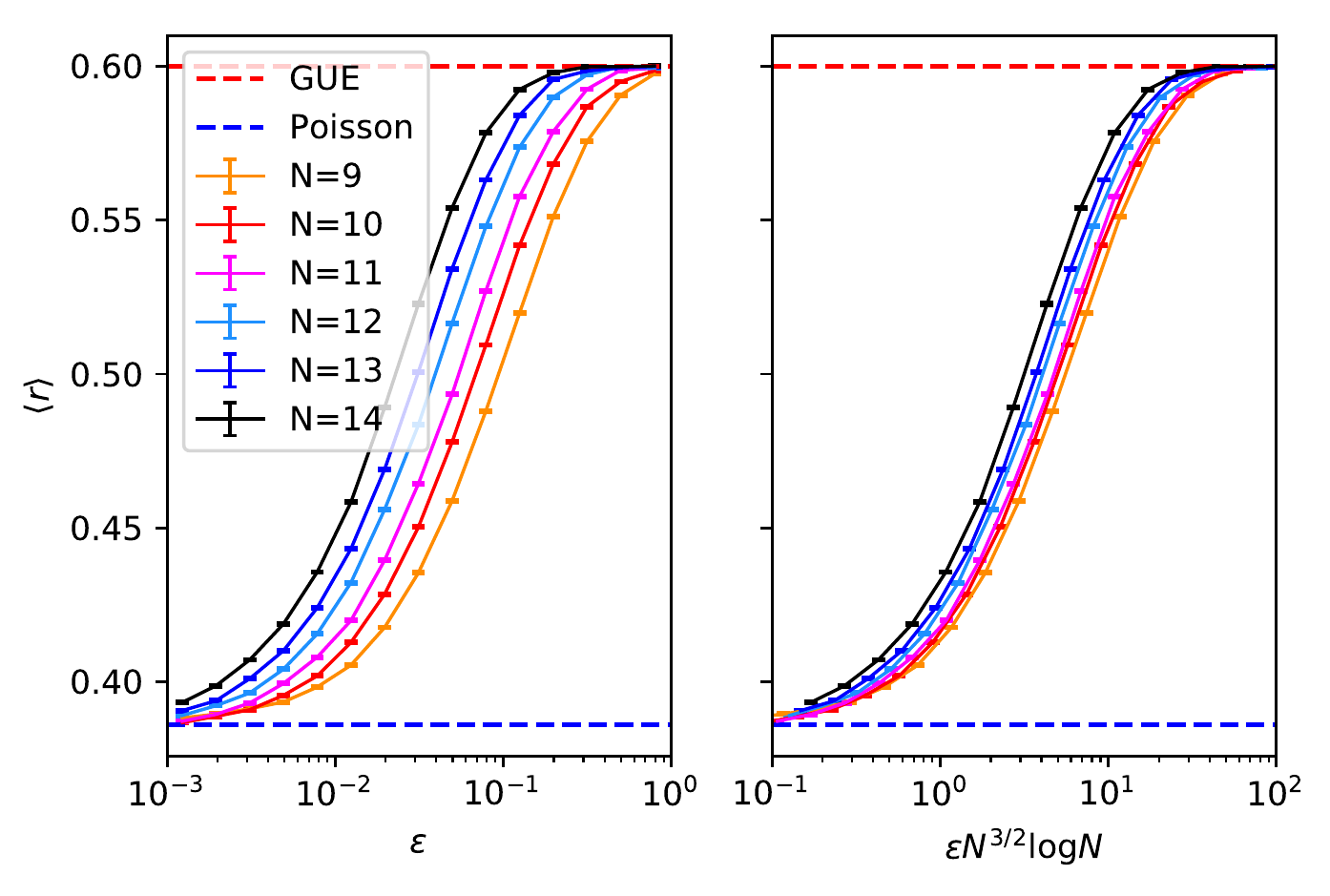}
    \includegraphics[width=0.7\linewidth]{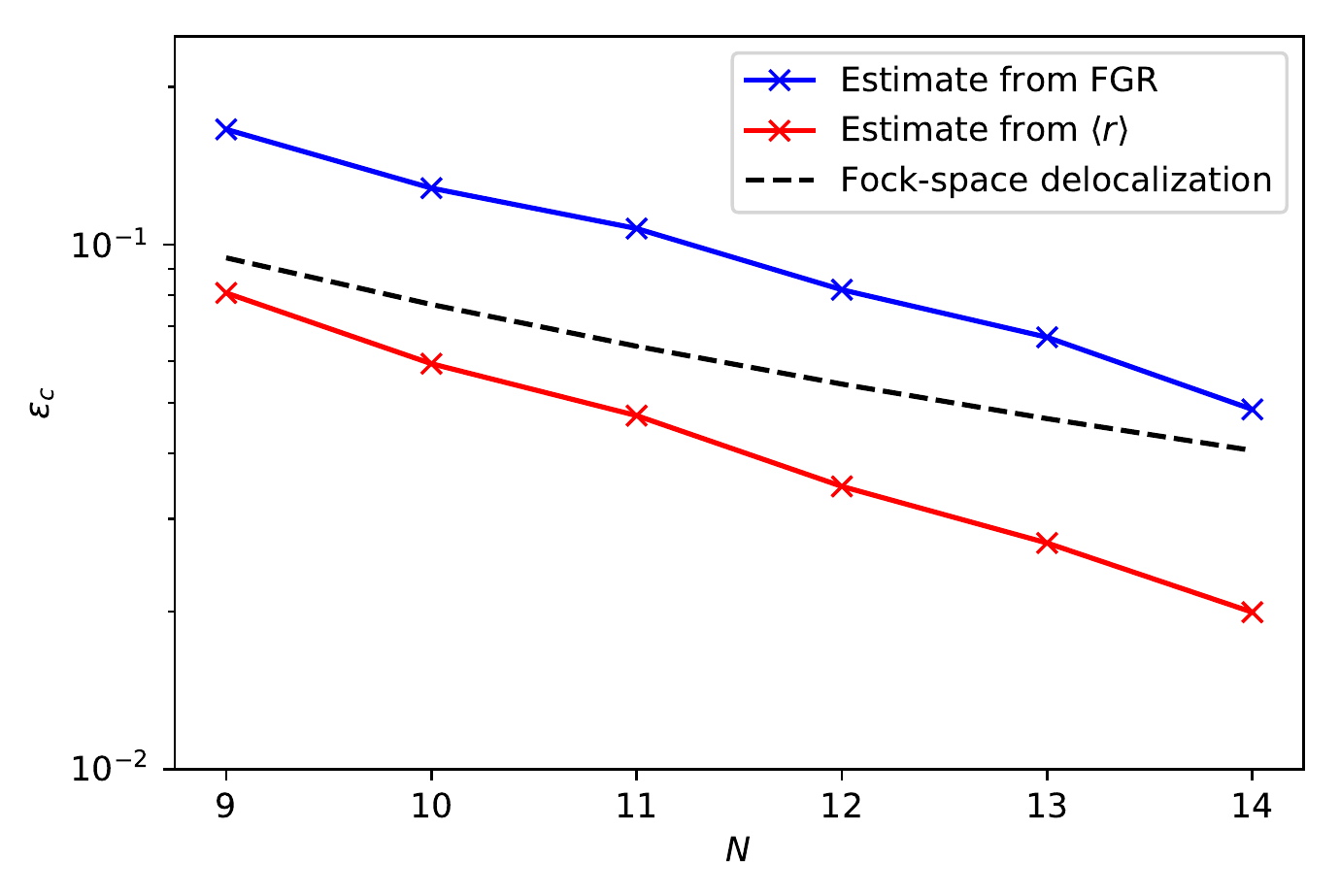}
    \caption{\textit{Top:} Scaling of $\langle r \rangle$ for the Hamiltonian Eq. \eqref{eq:GF}, according to theoretical expectations for a Fock space delocalization transition in this model Eq. \eqref{eq:Gaudin2scaling}. It is clear that thermalization sets in faster than expected for a Fock space delocalization transition. Simulation details as for Fig. \ref{Fig3}. Error bars denote the standard error of the sample mean over all realizations, which is $<0.1\%$ for all datapoints shown.
    \textit{Bottom:} Thresholds for the critical perturbation strength extracted from a Fermi golden rule estimate Eq. \eqref{eq:FGR} and the numerical $\langle r \rangle$ data respectively, according to the prescription described in Section \ref{sec:FGR}, for the model Eq. \eqref{eq:GF}. Black dotted lines are guides to the eye, and directly proportional to the Fock-space delocalization transition prediction for two-site hopping, Eq. \eqref{eq:Gaudin2scaling}. We see that both numerical estimates for $\epsilon_c$ decay notably faster than the delocalization transition estimate for accessible $N$.}
    \label{Fig8}
\end{figure}

We now consider what happens when the transverse Gaudin model is perturbed by operators that are non-local in the onsite spins, for example, a GUE two-fermion perturbation of the form
\begin{align}
\nonumber &\hat H_{\mathrm{Gaudin}+2F} = \hat H_0 + \epsilon \hat V, \quad \hat H_{0} = \frac{1}{\sqrt{N}} \sum_{i \neq j} J_i^* J_j \hat S_i^{+} \hat S_j^{-}, \\
\label{eq:GF}
&\hat V = \frac{1}{\sqrt{N}} \sum_{ij} t_{ij}\hat c_i^\dagger \hat c_j,
\end{align}
with $t_{ij} = t_{ji}^* = (X_{ij} + X_{ji}^*)/2$ and i.i.d. couplings $\mathrm{Re}(J_j)$, $\mathrm{Im}(J_j)$, $\mathrm{Re}[X_{ij}]$, $\mathrm{Im}[X_{ij}] \sim \mathcal{N}(0,1)$, and fermion operators defined via Jordan-Wigner strings
\begin{equation}
\hat c^\dagger_k = \prod_{j<k} (-\hat \sigma_j^z) \hat S^+_k.
\end{equation}
Na{\"i}vely, the two-fermion terms flip two spins and therefore correspond to a hopping distance $n_h=2$ in Fock space, with an associated critical value $\epsilon_c(N)$ for thermalization scaling as in Eq. \eqref{eq:Gaudin2scaling}. However, the attached Jordan-Wigner strings are highly non-local operators in the onsite basis, and generate long-range hopping in Fock space that leads to a delocalization crossover rather than a delocalization transition. Correspondingly, there is no scaling collapse to Eq. \eqref{eq:Gaudin2scaling} in $\langle r \rangle$ and we instead observe a drift towards thermalization, as depicted in Fig. \ref{Fig8}.

At the same time, it is difficult to rule out numerically the possibility that the long-ranged Fock space hopping in these models again results in a delocalization transition, but with a larger effective hopping range than the na{\"i}ve expectation $n_h=2$. Our numerical results are in fact consistent with a hopping range $n_h \approx 5$, but this hopping range does not have a natural microscopic interpretation that is consistent with the results of previous sections, suggesting that the agreement with this power-law fit might be spurious. We now turn to perturbed local integrable systems, for which the evidence for a crossover at exponentially small perturbation strengths is more stark.

\section{Perturbed local integrable systems}
\label{SecIV}
\subsection{Numerical analysis}

In this section we turn to spatially local integrable spin chains subject to spatially local perturbations. The example we will focus on is the XX spin chain with twisted boundary conditions, 
\begin{equation}\label{xxham}
\hat H_0 = \hat H_{\mathrm{XX}} = \sum_i \left[ e^{i \theta} \hat \sigma^+_i \hat \sigma^-_{i+1} + \mathrm{h.c.} + g \hat \sigma^z_i\right]. 
\end{equation}
The model in the absence of twisting and the uniform field has many degeneracies in its spectrum, related to $Z_2$ and reflection symmetry. Since we would like to have an integrable limit with Poisson level statistics, we fix $g = 0.3$, pick $\theta$ randomly for each sample, and limit our attention to odd system sizes $L$. (The last restriction eliminates some residual degeneracies in the half-filled sector; since we will consider a perturbation that breaks $U(1)$ conservation, the only way to avoid the half-filled sector is to work with $L$ odd.) 
 We choose 
\begin{equation}
\hat V = \sum_i (x_i \hat \sigma^x_i + y_i \hat\sigma^y_i) \hat \sigma^z_{i+1} ,
\end{equation}
where $x_i, y_i$ are i.i.d. random numbers chosen from the interval $(-1,1)$. This choice of $\hat H_0$ and $\hat V$ is motivated by the following logic. $\hat H_{\mathrm{XX}}$ maps to free fermions under a Jordan-Wigner transformation; this allows us to define an unambiguous Hamming distance in terms of the occupation numbers of the fermions, precisely as in AGKL. However, under the Jordan-Wigner mapping, our choice of $\hat V$ becomes nonlocal in fermion operators; thus, as in the generic interacting case, the interaction has matrix elements to arbitrarily large Hamming distance. Finally, since perturbations acting on adjacent sites do not commute, $\hat H$ in the limit of large $\epsilon$ becomes cleanly chaotic. If we had instead perturbed with random fields at every site, there would have been a further transition to a localized phase with increasing $\epsilon$, complicating our analysis of the onset of chaos.

\begin{figure}[tb]
\begin{center}
\includegraphics[width=0.45\textwidth]{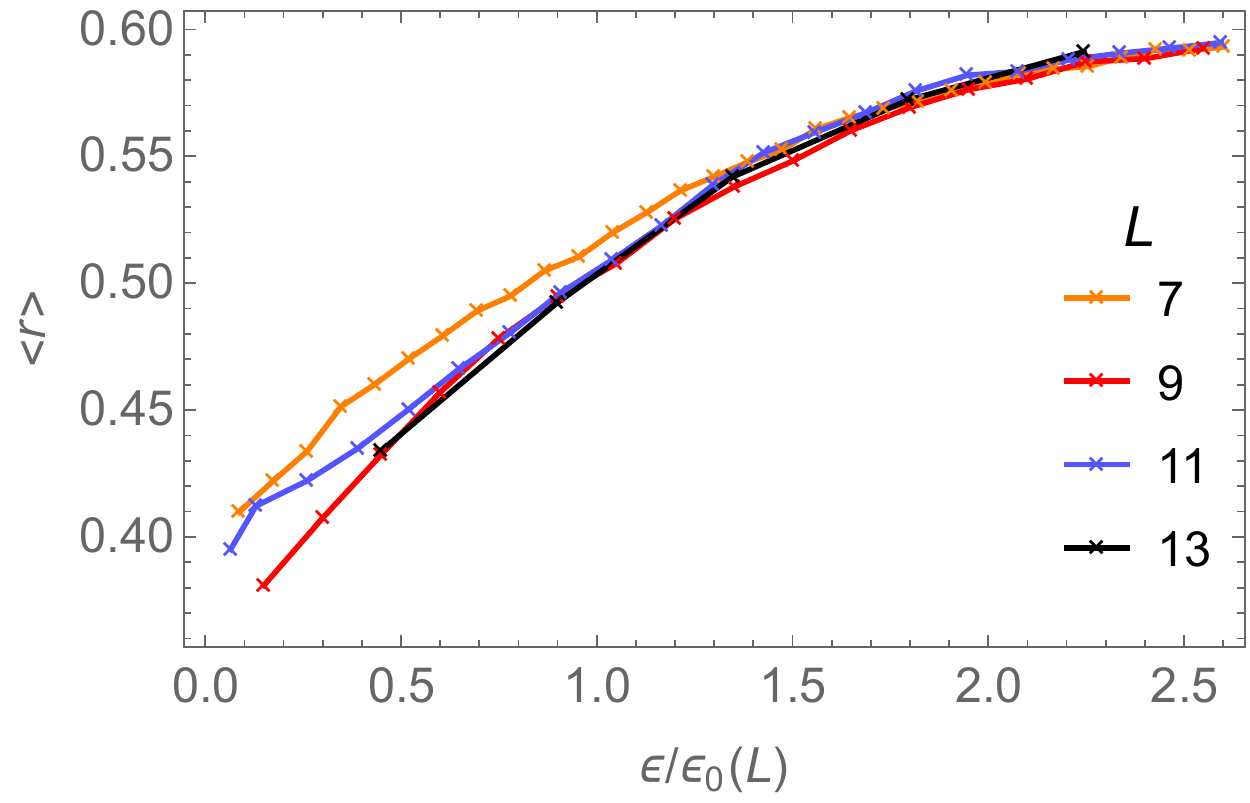}
\caption{Adjacent-gap ratio $\langle r \rangle$ vs. rescaled coupling $\epsilon/\epsilon_0(L)$, for the perturbed XX model, where $\epsilon_0(L) = C/1.32^L$, and $C$ is chosen so that $\langle r \rangle = 0.5$ when $\epsilon = \epsilon_0(L)$ for $L = 9$. The data are averaged over $5000$ samples for $L = 7, 9$; over $500$ samples for $L = 11$; and over $50$ samples for $L = 13$.}
\label{rcollapse}
\end{center}
\end{figure}

\begin{figure}[b]
\begin{center}
\includegraphics[width=0.45\textwidth]{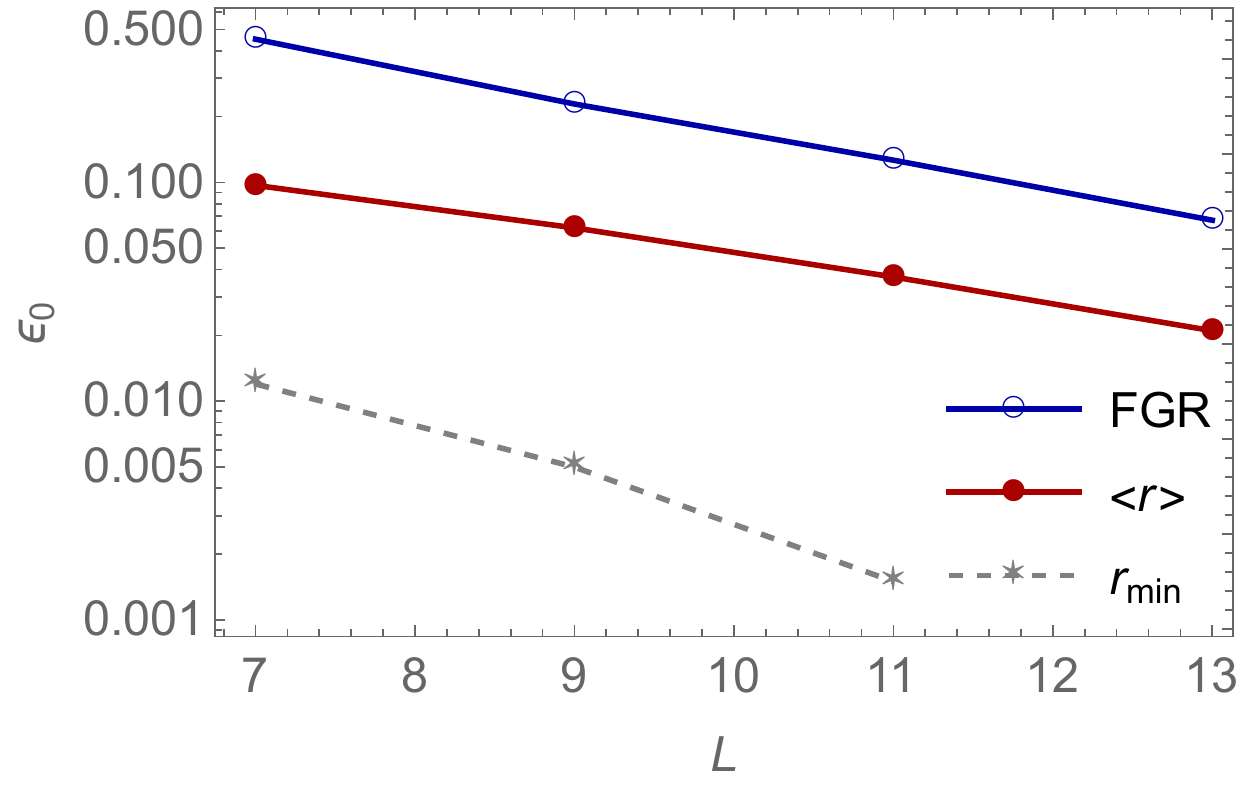}
\caption{Solid lines are two independent estimates of the crossover scale $\epsilon_0$, from Fermi's Golden Rule and the crossover in $\langle r \rangle$ respectively. Dashed lines estimate the crossover in the minimum gap $r_{\mathrm{min}}$, which happens at a parametrically smaller value of $\epsilon$.}
\label{xx_plot2}
\end{center}
\end{figure}

Our numerical analysis of the crossover to chaos in this model closely parallels that in the previous section. Fig.~\ref{rcollapse} shows the crossover in $\langle r \rangle$, which collapses when rescaled by a factor that is exponential in $L$. That the crossover does not sharpen appreciably with system size, and that the crossover scale varies exponentially with system size, are both suggestive evidence for a crossover rather than a transition in this model. Fig.~\ref{xx_plot2} shows the dependence of the crossover scale on system size, extracted from the Golden-Rule approach outlined in Sec. \ref{sec:FGR} as well as direct analysis of the crossover in $\langle r \rangle$. The methods agree reasonably well on the scaling with $L$. Although we cannot definitively tell an exponential from a power law with the modest dynamic range that is numerically accessible, fitting the data to a power law gives $\epsilon_0 \sim 1/L^\alpha$ with $\alpha > 3$, which has no microscopic rationale (since $\hat V$ induces transitions at Hamming distance 1). Finally, as Fig.~\ref{xx_plot2} shows, the crossover in the minimum gap occurs at parametrically lower values of $\epsilon$ than the crossover in $\langle r \rangle$, i.e., the regime~$(b)$ in Fig.~\ref{schematic} grows with system size~\footnote{For $L = 9$ there are some residual degeneracies in the integrable limit of the noninteracting model. These are too rare to affect $\langle r\rangle$ but do occur with high probability in typical samples. To lift these degeneracies, in our analysis of the minimum gap we have introduced an integrability-preserving interaction $0.05 \sum_i \hat \sigma^z_i \hat \sigma^z_{i+1}$. We have checked that the conclusions of the rest of this section are insensitive to the presence of this perturbation.}.

\subsection{Matrix elements of the Jordan-Wigner string}

\begin{figure*}[tb]
\begin{center}
\includegraphics[width = 0.95\textwidth]{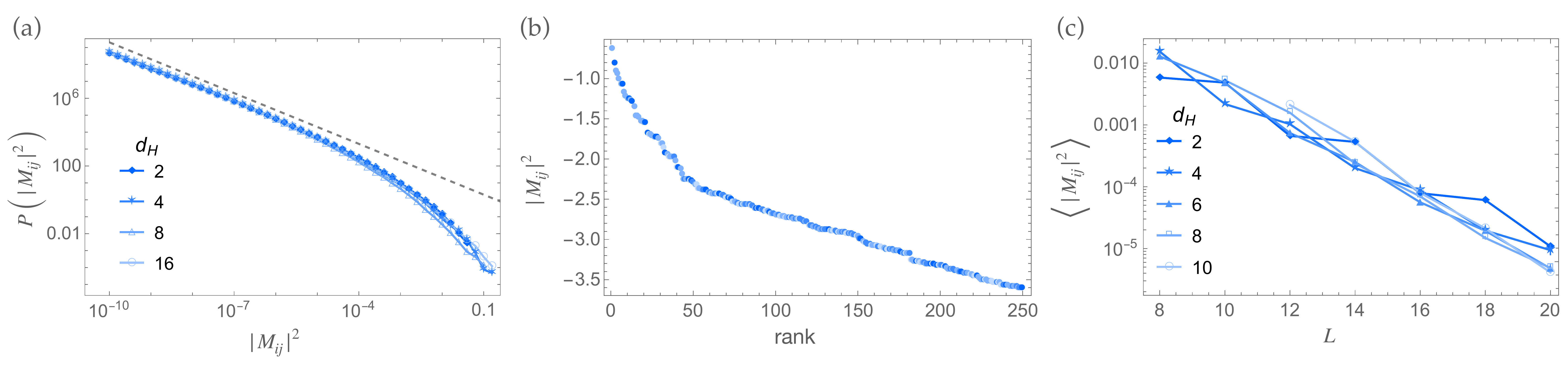}
\caption{(a)~Probability distribution of squared matrix elements $|M_{ij}|^2$ of the Jordan-Wigner string $\exp\left[i\pi\sum_{n < L/2}(1+\hat \sigma^z_n)/2\right]$, for system size $L = 20$, as a function of the Hamming distance $d_H$ between pairs of states $i, j$. The dashed line shows $P(x) \propto 1/x$ and corresponds to the matrix element distribution $P(|M_{ij}|) \propto 1/|M_{ij}|$. Data averaged over $2 \times 10^{6}$ pairs of states. (b)~Largest $250$ matrix elements for $L = 10$ (from an exhaustive list), color-coded by $d_H$. Lighter colors indicate larger $d_H$. (c)~Average matrix element $\langle |M_{ij}|^2 \rangle$ vs. $L$ for various Hamming distances, showing a clear exponential decay.}
\label{jwstring}
\end{center}
\end{figure*}

The rationale for our choice of the Hamiltonian~\eqref{xxham} was that for this model the unperturbed problem can be written in terms of free fermions, so one can define a natural Hamming distance between two states in terms of the difference in their occupation numbers. Each term in the perturbation $\hat V$---and generic parity-breaking perturbations---can be written in the form $\hat W(x) \hat O(x)$, where $\hat W(x) = \exp\left[i \pi \sum_{n<x} (1 + \hat\sigma^z_n)/2\right]$, and $\hat O(x)$ is a local operator that only rearranges a small number of fermions (and therefore, on its own, would only give transitions to some bounded Hamming distance). Thus, the large-scale rearrangements due to $\hat V$ are due to the matrix elements of $\hat W(x)$, which we now discuss. For this exploration, we choose open boundary conditions and work with initial states in which the fermions are at half filling; we also choose $x = L/2$. The matrix elements of $\hat W(x)$ between any pair of many-body Slater-determinant eigenstates can then be efficiently evaluated as a determinant~\cite{PhysRevLett.91.266602}.

Our results are shown in Fig.~\ref{jwstring}. We highlight a few key features. First, the natural Hamming distance is \emph{not} predictive of the strength of transitions: some of the largest matrix elements are to Hamming distances larger than 1. Intuitively, the Jordan-Wigner string looks like a pulsed step-function potential, and thus has larger matrix elements between states connected by small momentum transfer: thus, not all transitions at a given naive Hamming distance are equivalent. 
Second, the tail of the distribution to small matrix elements at any fixed Hamming distance has the form $P(|M_{ij}|) \sim 1/|M_{ij}|$. In the taxonomy of Sec.~\ref{fgrss} this tail has too little weight to provide the dominant transitions. At larger $|M_{ij}|$ the distribution bends and the apparent exponent increases to $P(|M_{ij}|) \sim 1/|M_{ij}|^2$, which is exactly the marginal case where transitions at all scales contribute to delocalization. Identifying whether a transition or crossover occurs will require a more intensive numerical study of (a)~the precise form of the tail to large matrix elements, and (b)~how the crossover between the two regimes of Fig.~\ref{jwstring}(a) scales with system size.
%
%
Third, the mean matrix element at any Hamming distance decays exponentially with system size with approximately the same decay constant, so that (entropically) the spectral weight for larger system sizes is dominated by transitions to large Hamming distance.

\section{Conclusion}

We have studied the system-size dependence of the onset of many-body quantum chaos in a variety of perturbed integrable systems. We have further argued that generalizations of the analysis of AGKL predicting a Fock-space delocalization transition for perturbed, non-interacting quantum dots apply to these systems.  Depending on the effective hopping range of the perturbation in the many-body Fock space defined by the eigenstates of the unperturbed integrable system, this analysis predicts either a crossover or a transition, at a perturbation strength that tends to zero as the system size $N \to \infty$. We have presented detailed numerical evidence that both possible scenarios are realized for perturbations of interacting or non-interacting integrable systems.

It is natural to ask whether analogous transitions can happen if one scales the integrability-breaking perturbation $\epsilon$ with \emph{time}, in an infinite system, rather than with system size. (This question was considered for MBL transitions in Ref.~\cite{GopalakrishnanHuse}.) For extended infinite integrable systems, the crossovers with time are much simpler than those with system size: thermalization in finite-size systems is slowed down by the discreteness of energy levels, which is not relevant for infinite systems.  The finite-time crossover occurs at a mean free time $\tau \sim 1/\epsilon^2$. The finite-time crossover in the dynamic limit of small $\epsilon$ can either happen through processes that primarily rearrange an order one number of quasiparticles, or through processes that rearrange a number of quasiparticles that diverges in this limit; this depends on whether the spectral weight of the operator $\hat V$ is dominated by small or large matrix elements. (In the power law model of Sec.~\ref{fgrss} this crossover from small-scale to large-scale rearrangements occurs at $\gamma = 3$, while the finite-size behavior changes its character at $\gamma = 2$.) The mean free time scales the same way with $\epsilon$ in both cases, but one might expect that the behavior of quantities such as out-of-time-order correlators (OTOCs) differs between these cases. Meanwhile, one experimentally relevant class of perturbed integrable systems where the temporal crossovers to chaos are not well understood, and might be elucidated by the considerations in this paper, are examples where integrability is broken by an external trapping potential\cite{cradle,hardrods,Calogero,PinToda1,PinToda2,bastianello2021hydrodynamics}. We leave further exploration of these questions for future work.


Our work raises at least two important questions for further theoretical analysis. The first is whether the statistics of the matrix elements of any perturbed interacting integrable systems can be characterized with sufficient accuracy to argue definitively for a Fock-space delocalization transition. The Gaudin-Richardson models studied numerically in this work seem to provide a natural starting point for tackling this question, since their eigenstates are particularly simple to write down explicitly and their couplings can be sampled randomly for any system size. The second question is whether the apparent simplicity of quantum integrability breaking in certain cases reflects a corresponding and hitherto unexplored simplicity in classical integrability breaking. For example, the sophisticated analysis of Wayne for interacting classical chains\cite{Wayne} seems to suggest a critical perturbation strength for breaking integrability that is polynomially small in the system size, as in Eq. \eqref{eq:Wayne}. This is intriguingly similar to the expected scaling of the Fock-space delocalization transition in some many-body quantum systems.

\section{Acknowledgments}
We thank A. Altland, D. Borgnia, X. Cao, A. Chandran, P. Crowley, V. Khemani, T. Micklitz, J. Moore, A. Morningstar, A. Polkovnikov, M. Rigol, D. Sels and R. Vasseur for stimulating discussions. S. G. is indebted to M. Rigol for pointing out an important typo in a draft of this paper. V.~B.~B. is supported by a fellowship at the Princeton Center for Theoretical Science. D.~A.~H. is supported in part by NSF QLCI grant OMA-2120757. S.~G. is supported in part by NSF grant DMR-1653271. The authors are pleased to acknowledge that the work reported on in this paper was substantially performed using the Princeton Research Computing resources at Princeton University which is consortium of groups led by the Princeton Institute for Computational Science and Engineering (PICSciE) and Office of Information Technology's Research Computing. 

\appendix

\section{Quantifying the Fock-space delocalization transition}
\label{Appendix}
\subsection{A greedy estimate for small denominators}
We now present a simple argument for how the power law dependence on $N$ in Eq. \eqref{eq:generalAGKL} can be understood as a consequence of small denominators in perturbation theory. This is closely related to the treatment of AGKL\cite{AGKL}, although it is too simple to capture the logarithmic correction; the connection with AGKL's results is discussed in the next section.

For concreteness, let $\hat H = \hat H_0 + \epsilon \hat V$. Consider $\hat H_0$ such that its eigenstates are product states over $N$ two-state spins, or fermionic Slater determinants for a fixed set of $N$ single-particle states.  In these cases the Fock space is simply represented as the corners of an hypercube.  Assume the hops in Fock space produced by $\hat V$ are of Hamming distance $n_h$.  Consider two eigenstates $|a\rangle$ and $|b\rangle$ of $\hat H_0$ that are adjacent in energy in the spectrum. Typically, these two states differ by Hamming distance $\cong N/2$. This means that the states $|a\rangle$ and $|b \rangle$ are first coupled at an order $l \cong N/(2n_h)$ in perturbation theory. Our goal is to estimate the effective coupling $V_{ab}^{\rm eff}$ between states $|a\rangle$ and $|b\rangle$ to leading order in the perturbation strength $\epsilon$.  It is this coupling that first (upon increasing $\epsilon$ from zero) produces hybridization and energy-level repulsion between $|a\rangle$ and $|b\rangle$ in the eigenstates and spectrum of $\hat H$.

First note that the leading coupling between $|a\rangle$ and $|b\rangle$ occurs at order $l$ in $\hat V$. The coupling is due to all ``paths'' $a\to a_1 \to \ldots a_{l-1} \to b$ when written in terms of the eigenstates of $\hat H_0$:
\begin{align}
\label{eq:denominators}
|V_{ab}^{\rm eff}|^2 = \epsilon^{2l}\sum_{a_1,a_2,\ldots,a_{l-1}} 
\frac{|V_{aa_1}|^2|V_{a_1a_2}|^2\ldots |V_{a_{l-1}b}|^2}{E_{aa_1}^2E_{aa_2}^2\ldots E_{aa_{l-1}}^2}
\end{align}
is the square of the leading-order coupling, where $E_{aa_n}=E_a-E_{a_n}$. We have chosen to expand around $|a\rangle$, but since $E_b$ is typically much closer to $E_a$ than any of the intermediate states, the expansion around $|b\rangle$ gives essentially the same result.

For non-interacting systems with delocalized quasiparticles, this sum is dominated by rare paths with the smallest energy denominators, since these paths give terms that are factorially (in $l$) larger than those from typical paths.  The size of this dominant contribution can be estimated from simple combinatorics, that is entirely analogous to arguments made in previous analyses of high-order perturbation theory\cite{AGKL,BAA}.

First, we obtain the number $N_{\mathrm{path}}(l)$ of shortest paths in Fock space connecting $|a\rangle$ to $|b\rangle$ via the hopping term $\hat V$ that are present in the sum in Eq. \eqref{eq:denominators}.  For simplicity, we consider the case where $|a\rangle$ and $|b\rangle$ differ by precisely $l n_h$ spin flips.

We may count the paths sequentially from $|a\rangle$ to $|b\rangle$:  Let $N_1$ denote the number of possible choices for $|a_1\rangle$, $N_2$ the number of possible choices for $|a_2\rangle$ having fixed $|a_1\rangle$, etc., until there is only one choice $N_l=1$ for $|a_l\rangle=|b\rangle$, having fixed $\{a_1,a_2,\ldots,a_{l-1}\}$. Then the total number of paths is given by $N_{\mathrm{path}}(l) = N_1 N_2 \ldots N_{l-1}$. 

Corresponding to this enumeration of shortest paths between $|a\rangle$ and $|b\rangle$ there exists a ``greedy'' algorithm for estimating the smallest path denominator $\Delta^*_{\mathrm{path}}$ in the sum Eq. \eqref{eq:denominators}. By Poisson level statistics of $\hat H_0$, once the state $|a_{k-1}\rangle$ is fixed, the ``next'' denominator $E_{aa_k}$ is approximately uniformly distributed over the typical bandwidth for $n_h$-particle hopping $w$. Since there are $N_k$ possible choices for $|a_k\rangle$, the average smallest denominator available to use at step $k$ is approximately
\begin{equation}
\Delta^*_k \approx \frac{w}{N_k}.
\end{equation}
Selecting the smallest denominator at each successive step leads to a greedy estimate
\begin{equation}
\label{eq:greedy}
\Delta^*_1 \Delta^*_2 \ldots \Delta^*_{l-1} \approx \frac{w^{l-1}}{N_1N_2\ldots N_{l-1}} = \frac{w^{l-1}}{N_{\mathrm{path}}}
\end{equation}
for the smallest path denominator contributing to the sum Eq. \eqref{eq:denominators}. Since greedy estimates are sub-optimal, this should bound the true smallest path denominator $\Delta^*_{\mathrm{path}}$ from above.  Each factor of $|V_{a_{k-1}a_k}/E_{aa_k}|$ in Eq. \eqref{eq:denominators} also has its numerator.  If the distributions of the matrix elements of $\hat V$ have a fat enough tail to very large values, then the largest term in the sum may also be set by that tail.  But for most choices of $\hat V$, including any choice where $\hat V$ has poly($N$) random coupling constants, these distributions will not have fat tails and the largest factor will be well approximated by the smallest denominator.  This yields an approximate \emph{lower} bound on the leading effective coupling
\begin{equation}
\label{eq:lower_bound}
\langle |V_{ab}^{\rm eff}| \rangle \geq \frac{t^{l}}{\Delta^*_{\mathrm{path}}} \gtrsim \frac{\epsilon (\epsilon/w)^{l-1} N_{\mathrm{path}}}{N^{l(n_h-1)/2}}~.
\end{equation}

Notice that this is the ``opposite'' of the classical KAM argument, which uses Diophantine conditions to bound small denominators in the perturbation series from below, thereby bounding the perturbation series from above and guaranteeing its convergence and the persistence of integrability\cite{poschel}. Here our strategy is instead to bound small denominators from \textit{above}, which bounds the perturbation series from below and yields an estimate for the onset of level repulsion and quantum chaos.

Our rough estimate of the onset of chaos is when $V_{ab}^{\rm eff}$ becomes of order the many-body level spacing $\sim w2^{-N}$.  Assuming that $n_h$ is of order one, and dropping order one factors, this yields
\begin{equation}
\label{eq:int_breaking_condition}
\epsilon_c \sim w N^{(n_h-1)/2} N_{\mathrm{path}}(l)^{-1/l}~.
\end{equation}
To relate this to Eq. \eqref{eq:generalAGKL}, note that the typical number of shortest paths scales as
\begin{align}
\nonumber N_{\mathrm{path}} &\sim \begin{pmatrix} N/2 \\ n_h \end{pmatrix} \begin{pmatrix} N/2-n_h \\ n_h \end{pmatrix} \ldots \begin{pmatrix} n_h \\ n_h \end{pmatrix} \\
\label{eq:Npath}
&\sim \left(\frac{N}{2e (n_h!)^{1/n_h}}\right)^{N/2}, \quad N \to \infty,
\end{align}
by Stirling's formula, which implies that
\begin{equation}
\label{eq:Anderson}
\epsilon_c(N) \sim \frac{w}{N^{(n_h+1)/2}}.
\end{equation}
This recovers Eq. \eqref{eq:ACAT} up to order one factors and the logarithmic correction (it in fact also corresponds to the Anderson criterion for delocalization\cite{AGKL}, i.e. hopping amplitude equals local level spacing). This argument also justifies why the $\langle r \rangle$ statistic, which measures level repulsion, should exhibit the same leading $N$-dependence as the threshold for Fock-space delocalization.

\subsection{AGKL-type estimates for the onset of many-body resonances}
We now illustrate how the above argument relates to the earlier analysis of AGKL\cite{AGKL}. In the language of our Fig. \ref{schematic}, the AGKL argument captures the transition from regime (b) to regime (c), at which typical many-body eigenstates become resonant with states that are adjacent in energy in the spectrum and Fock-space delocalization occurs. In fact, the AGKL analysis can be extended to model the crossover from regime (a) to regime (b), i.e. the onset of the first many-body resonances in the system, as we will discuss below.

Our starting point is the leading-order perturbative expression Eq. \eqref{eq:denominators} for the effective coupling $V_{ab}^{\mathrm{eff}}$ between two typical states that are adjacent in energy, which can be estimated in terms of the effective Fock-space hopping strength $t$ and many-body bandwidth $w$ as
\begin{equation}
|V^{\mathrm{eff}}_{ab}|^2 \approx \sum_{\vec{a}} \left(w \left(\frac{t}{w}\right)^l \prod_{j=1}^{l-1} \left| \frac{w}{E_{aa_j}}\right|\right)^2
\end{equation}
where the sum is over all lowest-order Fock space paths from $|a\rangle$ to $|b\rangle$. Let us introduce the notation
\begin{equation}
|V_{\vec{a}}| = w \left( \frac{t}{w}\right)^l \prod_{j=1}^{l-1} \left| \frac{w}{E_{aa_j}} \right|
\end{equation}
for each summand. Following AGKL, we model the $|V_{\vec{a}}|$ as i.i.d. random variables, and the energy denominators $E_{aa_j}$ as being i.i.d. uniformly distributed within the interval $[-w,w]$. This is of course an idealization that neglects the correlations between states, but is a reasonable assumption for typical states of an integrable system at large $N \gg 1$.

As noted by AGKL, it follows from these assumptions that $|V_{\vec{a}}| = w \left( \frac{t}{w}\right)^l e^{Y_{l-1}}$, with $Y_{l-1}$ a gamma distributed random variable with probability density function $f_{Y_{l-1}}(y) = y^{l-2}e^{-y}/\Gamma(l-1)$. Thus the cumulative distribution function of $|V_{\vec{a}}|$ is given by
\begin{equation}
\mathbb{P}(|V_{\vec{a}}| \geq U) = \int_{\log{[(U/w)(w/t)^l]}}^\infty \frac{y^{l-2}e^{-y}}{\Gamma(l-1)} dy,
\end{equation}
which implies the probability density function
\begin{equation}
f_{|V_{\vec{a}}|}(U) =  \frac{w}{\Gamma(l-1)U^2} \left(\frac{t}{w}\right)^l \log{\left[\frac{U}{w}\left(\frac{w}{t}\right)^l\right]^{l-2}}.
\end{equation}
Deep in the localized regime, $t \ll w$, the probability for no resonances of order $U$ among any of the $N_{\mathrm{path}}$ terms contributing to $V_{ab}^{\mathrm{eff}}$ can be written as
\begin{equation}
\label{eq:pnores}
(1-\mathbb{P}(|V_{\vec{a}}| \geq U))^{N_{\mathrm{path}}} = e^{-x_N(U)}
\end{equation}
where $x_N(U) \approx N_{\mathrm{path}} \mathbb{P}(|V_{\vec{a}}| \geq U)$. To proceed further, it is useful to note the approximation
\begin{equation}
\label{eq:papprox}
\mathbb{P}(|V_{\vec{a}}| \geq U) \approx \frac{w}{(l-2)!} \frac{1}{U\log^2{\left[\frac{U}{w}\left(\frac{w}{t}\right)^l\right]}} \left(\frac{t}{w}\log{\left[\frac{U}{w}\left(\frac{w}{t}\right)^l\right]}\right)^l
\end{equation}
valid for $U = \mathcal{O}(1)$. Now by Eq. \eqref{eq:pnores} the threshold for a typical state to become resonant with its nearest neighbour in energy is given by
\begin{equation}
\label{eq:bccondition}
N_{\mathrm{path}} \mathbb{P}(|V_{\vec{a}}| \geq w2^{-N}) \approx 1.
\end{equation}
To simplify this expression, first note that by Eq. \eqref{eq:papprox} and Stirling's formula,
\begin{equation}
\mathbb{P}(|V_{\vec{a}}| \geq w2^{-N}) \approx \frac{1}{\sqrt{2\pi l}} \frac{1}{\log^2{\left[\frac{w}{2^{2n_h}t}\right]}} \left[\frac{e2^{2n_h}t}{w} \log{\frac{w}{2^{2n_h}t}} \right]^l.
\end{equation}
From Eq. \eqref{eq:Npath}, it follows that for large $N$ (equivalently large $l$) the left hand side of Eq. \eqref{eq:bccondition} is dominated by the contribution exponential in $N$:
\begin{equation}
x_N(w2^{-N}) \sim \left[\frac{2^{n_h}}{e^{n_h-1} n_h!} \frac{zt}{w} \log{\left(\frac{w}{t2^{2nh}}\right)}\right]^l,
\end{equation}
which implies that the threshold Eq. \eqref{eq:bccondition} sets in when
\begin{equation}
\label{eq:AGKLresonance}
\frac{w}{t} \approx C z\log \frac{w}{t},
\end{equation}
where the numerical prefactor $C = 2^{n_h}/(e^{n_h-1}n_h!)$. This self-consistently yields the AGKL-like criterion
\begin{equation}
\frac{w}{t} \sim C z \log{z} + \mathcal{O}(z).
\end{equation}

One could instead ask for the threshold at which the first many-body resonances occur between pairs of states. This will in general depend on the typical Hamming distance $\alpha N$ between the pairs of states under consideration. So far we have been considering typical states that are adjacent in the energy spectrum, for which $\alpha=1/2$ sets the expected Hamming distance as $N \to \infty$. Let us now relax this assumption and consider arbitrary values $0<\alpha<1$. We will find that the threshold for the first resonance at typical Hamming distance $\alpha N$ can depend strongly on $\alpha$.

First consider the effective matrix element $|V_{ab}|^2$ for states $|a\rangle$ and $|b\rangle$ at a Hamming distance $\alpha N$. The lowest order process coupling these states is now at order $l= \alpha N/n_h$. Thus the number of shortest Fock space paths connecting $|a\rangle$ to $|b\rangle$ now depends on $\alpha$, and is given by
\begin{align}
\nonumber N_{\mathrm{path}} &\sim \begin{pmatrix} \alpha N \\ n_h \end{pmatrix} \begin{pmatrix} \alpha N -n_h \\ n_h \end{pmatrix} \ldots \begin{pmatrix} n_h \\ n_h \end{pmatrix} \\
&\sim \left(\frac{\alpha N}{e(n_h!)^{1/n_h}}\right)^{\alpha N}, \quad N \to \infty.
\end{align}

Next note that for a given state $| a \rangle$, there are $d_\alpha = \begin{pmatrix} N \\ \alpha N \end{pmatrix}$ other states at a Hamming distance $\alpha N$ from $|a\rangle$. By Poisson statistics, the state among these closest to $|a\rangle$ in energy will typically have an energy gap $|E_{ab}| \sim w/d_\alpha$ relative to $|a\rangle$. By Eq. \eqref{eq:pnores}, the probability that any given eigenstate $|a \rangle$ is resonant with its nearest energy neighbour $|b\rangle$ at Hamming distance $\sim \alpha N$ is set by the condition
\begin{equation}
N_{\mathrm{path}} \mathbb{P}(|V_{\vec{a}}| \geq w/d_\alpha) \approx 1.
\end{equation}
Introducing the constant $c_{\alpha}$ such that $d_{\alpha} = c_{\alpha}^l$, we find that
\begin{equation}
\mathbb{P}(|V_{\vec{a}}| \geq w/d_{\alpha}) \approx \frac{1}{\sqrt{2\pi l}} \frac{1}{\log^2{\left[\frac{w}{c_{\alpha}t}\right]}} \left[\frac{e c_{\alpha}t}{w} \log{\frac{w}{c_{\alpha}t}} \right]^l.
\end{equation}
It follows that in this case, the dominant contribution to $x_N$ at large $N$ is given by
\begin{equation}
\label{eq:xna}
x_N(w/d_\alpha) \sim \left[\frac{\alpha^{n_h}c_\alpha}{e^{n_h-1} n_h!} \frac{z t}{w} \log \left(\frac{w}{c_{\alpha}t}\right)\right]^l.
\end{equation}
Thus typical states experience resonances at Hamming distance $\sim \alpha N$ when
\begin{equation}
\frac{w}{t} \sim C_{\alpha} z \log z + \mathcal{O}(z),
\end{equation}
where the $\mathcal{O}(1)$ constant 
\begin{equation}
C_{\alpha} = \frac{c_\alpha \alpha^{n_h}}{e^{n_h-1}n_h!} \sim \frac{1}{(1-\alpha)^{n_h(\alpha^{-1}-1)}e^{n_h-1}n_h!}, \quad N \to \infty.
\end{equation}

From here, one can estimate a threshold for the \emph{first} resonances at Hamming distance $\alpha N$. Neglecting correlations between states, as in AGKL, it follows from Eq. \eqref{eq:pnores} that the probability for at least one resonance at Hamming distance $\alpha N$ among all eigenstates $|a \rangle$ is set by the condition
\begin{equation}
2^N x_N(w/d_\alpha) \approx 1.
\end{equation}
By Eq. \eqref{eq:xna}, this happens at the threshold
\begin{equation}
\frac{w}{t} \sim 2^{n_h/\alpha}C_{\alpha} z \log z + \mathcal{O}(z),
\end{equation}
which amounts to an upwards renormalization of the ``effective coordination number'' by a factor $2^{n_h/\alpha}$. For pairs of states that are neighbouring in energy, for which typically $\alpha \approx 1/2$, this upwards renormalization is by a factor $z \mapsto 2^{2n_h} z$.

We deduce that at any given Hamming distance scale $\sim \alpha N$, the first resonances at this Hamming distance (corresponding to one definition for the threshold between regimes (a) and (b) in Fig. \ref{schematic}) happen at a weaker perturbation strength than the corresponding transition between regimes (b) and (c), as one might expect. However, the much stronger $\alpha$-dependence (e.g. as $\alpha \to 0$) of the threshold for ``first resonances" than for ``typical resonances" indicates that the distinction between (a) and (b) is fuzzier than that between (b) and (c), and therefore more properly viewed as a crossover than as a transition.
\bibliography{intbib.bib}
\end{document}